\documentclass[twoside,11pt]{article}

\usepackage{jmlr2e}

\usepackage{amsmath}
\usepackage{graphicx}
\usepackage{natbib}
\usepackage{url} 
\usepackage{smile}

\usepackage{hyperref}       
\usepackage{url}            
\usepackage{booktabs}       
\usepackage{amsfonts}       
\usepackage{nicefrac}       
\usepackage{microtype}      
\usepackage{xcolor}         
\usepackage{smile}
\usepackage{amsmath,amssymb}
\usepackage{psfrag,epsf}
\usepackage{enumerate}
\usepackage{natbib}
\usepackage{algorithm}
\usepackage{algorithmic}
\usepackage{bm}
\usepackage{mathtools}
\usepackage{wrapfig}
\usepackage{lipsum}
\usepackage{mathrsfs}
\usepackage{dsfont}
\usepackage{rotating}
\usepackage{float}
\usepackage{natbib}
\usepackage{multirow}
\usepackage{apalike}
\usepackage{subfig}

\def\reals{{\mathbb R}}
\def\P{{\mathbb P}}
\def\Pc{{\mathcal P}}

\def\M{{\mathcal M}}

\def\Cb{{\mathbb C}}
\def\D{{\mathbb D}}
\def\vec{{\mathop{\text{\rm vec}}}}

\def\E{{\mathbb E}}
\def\R{{\mathcal{R}}}

\usepackage{lastpage}
\jmlrheading{23}{2022}{1-\pageref{LastPage}}{1/21; Revised 5/22}{9/22}{21-0000}{Zeyu Li, et al.}

\ShortHeadings{Simultaneous Estimation and Useful Dataset Selection}{Zeyu Li, et al.}
\firstpageno{1}

\begin{document}

\title{Simultaneous Estimation and Dataset Selection for Transfer Learning in High Dimensions by a Non-convex Penalty}

\author{\name Zeyu Li  \email zeyuli21@m.fudan.edu.cn \\
       \addr School of Management\\
       Fudan University\\
       \AND
       \name Dong Liu \email liudong\_stat@163.com \\
       \addr School of Physical and Mathematical Sciences\\
       Nanyang Technological University\\
        \AND
       \name Yong He  \email heyong@sdu.edu.cn \\
       \addr Institute for Financial Studies\\
       Shandong University\\
        \AND
       \name Xinsheng Zhang \email xszhang@fudan.edu.cn \\
       \addr School of Management\\
       Fudan University
       }

\footnotetext[1]{The first two authors contributed equally to this work. Yong He is the corresponding author.}

\editor{My editor}

\maketitle

\begin{abstract}
In this paper, we propose to estimate model parameters and  identify informative source datasets simultaneously for high-dimensional transfer learning problems with the aid of a non-convex penalty, in contrast to the separate useful dataset selection and transfer learning procedures in the existing literature. To numerically solve the non-convex problem with respect to two specific statistical models, namely the sparse linear regression and the generalized low-rank trace regression models, we adopt the difference of convex (DC) programming with the alternating direction method of multipliers (ADMM) procedures. We theoretically justify the proposed algorithm from both statistical and computational perspectives. Extensive numerical results are reported alongside to validate the theoretical assertions. An \texttt{R} package \texttt{MHDTL}\footnote[2]{The \texttt{R} package is available at \url{https://github.com/heyongstat/MHDTL}.} is developed to implement the proposed methods.
\end{abstract}


\section{Introduction}

\label{sec:intro}

The idea of transfer learning, originated from the computer science community \citep{torrey2010transfer,zhuang2020comprehensive,niu2020decade}, has been applied on various \emph{high-dimensional} statistical problems in the past few years \citep{bastani2021predicting,li2020transfer,tian2022transfer,cai2021transfer}. As its name indicates, useful information from related tasks (\emph{sources}) could be transferred to the original task (\emph{target}), so as to improve the efficiency of statistical inference for the latter.

Under the high-dimensional statistics setting where the number of parameters can be much larger than the sample sizes, additional low-dimensional structures are often imposed on the model parameters, in order to  avoid curse of dimensionality and derive  consistent estimators. With the intrinsic low-dimensionality assumption, flourishing regularized (or penalized) convex optimization methods have been proposed to achieve statistically optimal performance within polynomial time. One typical example is the sparse signal assumption, i.e.,  assuming the sparsity of model parameters of interest \citep{tibshirani1996regression,fan2001variable,candes2007dantzig}, which often involves the lasso type penalty. When the parameters of interest arise in the matrix form, an alternative model assumption would be low-rankness, which has been widely explored and applied in the communities of statistics, computer science and econometrics \citep{zhou2014regularized,fan2019generalized,He2021Vector}. In this article, we follow \cite{negahban2012unified} and work under the high-dimensional M-estimator framework, which includes sparse linear regression and generalized low-rank trace regression models as concrete examples.

\subsection{Literature review and contributions}

Intuitively, when the sources are sufficiently similar to the target, transfer learning could outperform the direct estimation procedure using only the target dataset. Specifically, let's consider the ideal case when all source distributions are identical to the target distribution, then the statistically optimal estimation procedure is to first pool all the datasets and make a global pooling estimation. However, the knowledge transfer estimators might sometimes have much poorer performance than the single-task learning using the target dataset alone, also known as \emph{negative transfer} \citep{torrey2010transfer}. This naturally leads to the following question:  \emph{what might be the reason of  negative transfer in the high-dimensional statistical context?} 

The fundamental reason is that the data distributions of the target and sources are generally not identical. It is then typically summarized in the following two points in the high-dimensional supervised transfer learning literature \citep{he2024transfusion}: 1. \emph{model shift} which  often refers to the difference in model parameters, e.g., the regression coefficients; 2. \emph{covariate shift} which is often related to the Hessian matrices of the population loss functions, e.g., the population covariance matrices of covariates in the case of linear regression.

In \cite{li2020transfer,tian2022transfer}, the authors quantify the model shift using certain distance measure of the target and source parameters. We also adopt the distance similarity in this article: if the source parameter vector is sufficiently close to the target one, we say the source is useful (or informative) to the target task, or the model shift is slight. When the informative sources are known in advance, which is also called the oracle transfer setting in the literature, \cite{bastani2021predicting,li2020transfer,tian2022transfer} propose to use a two-step procedure: in the first step an oracle pooling estimator is acquired by pooling the useful datasets, and in the second step this estimator is then debiased using only the target data. Meanwhile, in the computer science literature, the idea of fitting a shared global model for all related local clients even appears in personalized federated learning, e.g., the \texttt{FEDAVG} algorithm \citep{mcmahan2017communication}, when the heterogeneity of data distributions is in mind. To some extent, \texttt{FEDAVG} targets at the oracle pooling estimator. Indeed, in Theorem 3.3 of \cite{chen2021theorem}, the authors show that the minimax optimal rate in knowledge transfer is achieved by either single-task learning or data pooling estimation under the distance similarity in low-dimensions.

In high-dimensional cases, the previously mentioned two-step methods are often sensitive to covariate shift, even if the model shift is slight for all sources. Hence, these methods often require additional homogeneous conditions on the Hessian matrices of the population loss functions. Quite recently, \cite{li2023estimation,he2024transfusion} give sophisticated knowledge transfer methods that take both slight model shift and covariate shift into account. Enlightened by the claims of \cite{chen2021theorem} concerning low-dimensions, we show that the statistical performance of the high-dimensional oracle pooling estimator is still guaranteed by simply enlarging the regularization parameter from the default rate, regardless of the covariate shift. Specific statistical models are discussed within our framework, including that of the generalized low-rank trace regression whose knowledge transfer problem, to the best of our knowledge, has not been discussed in the existing literature.

More unfortunately, the useful sources with small model shift are generally unknown in practice. That is to say, there might be useless and even potentially harmful sources with generally larger model shift included for consideration. To deal with this problem and acquire the oracle pooling estimator in the non-oracle setting, \cite{li2020transfer,li2023estimation} take advantage of model selection aggregation, while \cite{tian2022transfer} propose a data-driven method to first detect the transferable sources in a separate step before applying their oracle transfer learning algorithm. However, these methods often require solving optimization problems repeatedly with the entire source datasets, which results in additional computational burden and loss of data privacy. 

In this work, we introduce a non-convex penalty, which is originated from the clustering analysis literature \citep{shen2012likelihood,Pan2013Cluster, wu2016new,liu2023cluster}, to the transfer learning context. This enables automatically incorporating those highly informative sources (with slight model shift) into the estimation procedure of the target task, while ignoring the non-informative ones (with larger model shift). In contrast to the existing methods focusing on identifying useful dataset as a separate step, our algorithm directly outputs the estimator by simultaneously selecting the useful sources. Moreover, the DC-ADMM algorithm in this article only needs the parameters from the sources during the iterations, instead of the whole datasets. This is particularly important if preserving data privacy and taking care of communication cost are also in mind \citep{mcmahan2017communication}.

In summary, the contributions of this article are stated as follows. We propose to introduce a non-convex penalty to estimate the target parameters while simultaneously selecting the useful sources. A difference of convex (DC) programming with the alternating direction method of multipliers (ADMM) algorithm is used to solve the  non-convex optimization problem with respect to two specific statistical models, including generalized low-rank trace regression whose knowledge transfer problem is not discussed in the literature as far as we know. The proposed algorithm is theoretically justified from both statistical and computational aspects. An \texttt{R} package \texttt{MHDTL} is developed, which is also used in the extensive numerical experiments that support our claims.

\subsection{Framework}

We consider the transfer learning problem under the high-dimensional M-estimators framework with decomposable regularizers following \cite{negahban2012unified}. We denote the target distribution as $\D_0$ and $K$ potential source distributions as $\D_k$, $k=1,\cdots,K$, whose population parameters are denoted as $\btheta^*_k\in \reals^p$, $k\in\{0\}\cup[K]$. We assume that $\btheta^{*}_k$ is the minimizer of the expected loss function, i.e.,
\begin{equation*}
	\btheta^{*}_k = \argmin_{\btheta\in \reals^p} \E \ \cL_k(\bZ_k;\btheta), \quad \bZ_k \text{ is sampled from } \D_k,
\end{equation*}
where $\cL_k(\cdot;\btheta)$ is the loss function of the $k$-th study. In this work, we mainly focus on high-dimensional regression problems and $K$ is treated as a finite constant. For instance, we can let the $k$-th dataset consist of $n_k$ {\it i.i.d.} observations of $\bZ_{k,i} = (\bX_{k,i},y_{k,i})$, set $y_{k,i} = \langle \btheta^*_{k}, \bX_{k,i} \rangle +\varepsilon_{k,i}$, where $\varepsilon_{k,i}$ are standard Gaussian errors, then  $\btheta^*_k$ would be the minimizer of the expected loss function $\E (\|y_{k}-\langle \btheta, \bX_{k}\rangle\|^2)$. For another example, one may consider the generalized linear model $\P(y_{k,i}|\bX_{k,i}) \propto \exp\left\{y_{k,i} \langle \btheta^*_k, \bX_{k,i} \rangle-b( \langle \btheta^*_k, \bX_{k,i} \rangle) \right\}$, where $\btheta^*_k= \argmin_{\btheta\in \reals^p}\E \left\{-y_{k}\langle \btheta, \bX_{k} \rangle + b(\langle \btheta, \bX_{k} \rangle) \right\}$.

In the high-dimensional statistical context, we assume that $\btheta^*_0$ lies in a low-dimensional subspace $\M\subset \reals^p$. For instance, it might be the subspace of vectors with a particular support or the subspace of low-rank matrices. Denote the inner product induced norm as $\|\cdot\|^2=\langle \cdot,\cdot\rangle$. Given subspace $\overline{\M}$ such that $\M\subset \overline{\M}$, let $\overline{\M}^{\perp}=\{\bv\in \reals^p\,|\,\langle \bu,\bv\rangle=0 \text{ for all } \bu\in \overline{\M}\}$ be the orthogonal complement of the space $\overline{\M}$, we say a norm-based regularizer (or penalty) $\R$ is decomposable with respect to $(\M,\overline{\M}^{\perp})$ if 	
$$\R(\btheta+\bgamma)=\R(\btheta)+\R(\bgamma), \quad \text{for all } \btheta \in \M \text{ and } \bgamma\in \overline{\M}^{\perp}.$$
For the sake of better illustration, we present here several examples of the decomposable regularizers given in \cite{negahban2012unified}.

\begin{example}[Sparse vector and $\ell_1$ norm]
 	Suppose that $\btheta^*_0$ is supported on some subset $S\subset \{0,1,\dots,p\}$ with cardinality $s$, we set $\cM=\overline{\cM} =\{\btheta\in\reals^p\mid \btheta_j=0 \text{ for all } j\notin S \}$. We can take $\cR(\cdot)=\|\cdot\|_1$, it is easy to verify that $\|\btheta+\bgamma\|_1=\|\btheta\|_1+\|\bgamma\|_1$ for all $\btheta \in \M$ and $\bgamma\in \overline{\M}^{\perp}$ by the construction of the subspaces. If we consider the linear regression model, we face the transfer learning problem for sparse linear regression discussed in \cite{li2020transfer}. If we take the generalized linear model instead, the problem degenerates to the one discussed in \cite{tian2022transfer}.
 \end{example}

 \begin{example}[Low-rank matrix and nuclear norm]
 	Let $\btheta^*_0$ be a low-rank matrix, and let $\btheta^*_0=\bU\bD\bV^{\top}$ be its singular value decomposition (SVD), where $\bD$ is a diagonal matrix that consists of non-increasing singular values. Denote the first $r$ columns of $\bU$ and $\bV$ by $\bU^r$ and $\bV^r$, we take
 	\begin{equation}\label{eq:gltr1}
 	\btheta^*_0\in\mathcal{M}:=\left\{\btheta \in \mathbb{R}^{d_1 \times d_2} \mid \operatorname{row}(\btheta) \subseteq \operatorname{col}\left(\mathbf{V}^r\right), \operatorname{col}(\btheta) \subseteq \operatorname{col}\left(\mathbf{U}^r\right)\right\},
 	\end{equation}
 	\begin{equation}\label{eq:gltr2}
 		\overline{\mathcal{M}}^{\perp}:=\left\{\btheta \in \mathbb{R}^{d_1 \times d_2} \mid \operatorname{row}(\btheta) \perp \operatorname{col}\left(\mathbf{V}^r\right), \operatorname{col}(\btheta) \perp \operatorname{col}\left(\mathbf{U}^r\right)\right\},
 	\end{equation}
where $\operatorname{col}(\cdot)$ and $\operatorname{row}(\cdot)$ denote spaces spanned by columns and rows. It is known that the matrix nuclear norm $\|\cdot\|_{\text{N}}$ satisfies the decomposability condition. Note that here $\M$ is not equal to $\overline{\M}$. Similarly, given the linear regression model, we now have transfer learning problems for low-rank trace regression, which is introduced in \cite{zhou2014regularized} and is particularly useful in modeling matrix completion, multi-task learning and compressed sensing problems \citep{hamdi2022low}. Meanwhile, for the generalized linear model we have generalized low-rank trace regression with various applications including generalized reduced-rank regression and one-bit matrix completion \citep{fan2019generalized}.
 \end{example}

Finally, the characterization of similarity between the target and sources is essential for us to design algorithms that fully utilize the shared information and also to provide sound theoretical analysis. As previously alluded, we consider the following popular distance similarity: for the informative sources in a subset $\cA\subseteq \{1,\cdots K\}$, we control their magnitudes of model shift by giving an upper bound on the distance of the difference vectors. Namely, for $\bdelta^*_k=\btheta^*_0-\btheta^*_k$, we assume for some norm $\cB_k$ that
\begin{equation}\label{eq-ldc}
	\cB_k(\bdelta^*_k)\leq h,\quad k\in\cA,
\end{equation}
where smaller $h$ indicates higher similarity. For example, $\cB_k$ can be the vector $\ell_1$ or $\ell_2$ norm in the case of sparse linear regression. For generalized trace regression, it can be the nuclear norm, the Frobenius norm or the vectorized $\ell_1$ norm. As for $k\in \cA^c$, $\btheta^{*}_k$ is allowed to be quite different from $\btheta^*_0$. Meanwhile, we make no assumption on covariate similarity, i.e., there is no additional homogeneous condition on the Hessian matrices of the population loss functions as in \citep{li2020transfer,tian2022transfer}.

\subsection{Organization and notations}
The remainder of this paper is organized as follows. In Section \ref{sec:method}, we introduce the truncated norm penalty and the resulting non-convex optimization problem. We then give the computational details for solving the non-convex problem under the sparse linear regression and generalized low-rank trace regression settings. In Section \ref{sec:theory}, we discuss the statistical properties of the proposed algorithm. We report extensive numerical simulation results in Section \ref{sec:simulation}. Finally, the proposed algorithm is applied on two real datasets in Section \ref{sec:real}. In the appendix, we give additional numerical implementation details and also the proofs of the theoretical results.

To end this section, we state some commonly used notations. The inner product induced norm $\|\cdot\|$ is sometimes written as $\|\cdot\|_2$ for vectors or $\|\cdot\|_{\text{F}}$ for matrices. For real vector $\ba$, let $\|\ba\|_1$ be its $\ell_1$-norm, while for real matrix $\bA$, let $\|\bA\|_{\text{op}}$ and $\|\bA\|_{\text{N}}$ be its operator norm (from $\ell_2$ to $\ell_2$) and nuclear norm respectively. We use the standard $O_p$ notation for stochastic boundedness. For a sub-Gaussian random variable $X$, we define the sub-Gaussian norm by $\|X\|_{\Psi_2}=\inf \left\{t>0: \mathbb{E} \exp \left(X^2 / t^2\right) \leq 2\right\}$, while for a random vector $\bX$, its sub-Gaussian norm is defined as $\|\bX\|_{\Psi_2}=\sup _{\|\bx\|_2=1}\|\langle \bX, \bx\rangle\|_{\Psi_2}$. In the end, we write $x\lesssim y$ if $x\leq Cy$ for some $C>0$, $x\gtrsim y$ if $x\geq cy$ for some $c>0$, and $x\asymp y$ if both $x\lesssim y$ and $x\gtrsim y$ hold. Note that the constants $C$ and $c$ may not be identical in different lines.

\section{Methodology}\label{sec:method}

In the oracle case, we know which of the sources are informative, namely we know the subset $\cA$ of small model shift in advance. Let $\cP:=\{0\}\cup \cA$, for $n_{\cP}:=\sum_{k\in \Pc}n_k$, the oracle pooling estimator is
 	\begin{equation}\label{weighted-pooling}
			\hat{\btheta}_{\cP} =\argmin_{\btheta\in \reals^p} \frac{1}{n_{\cP}}\sum_{k\in \cP}\sum_{i\leq n_k} \cL_k(\bZ_{k,i};\btheta) + \lambda_{\cP} \R(\btheta).
	\end{equation}

 When the informative sources are unknown, to eliminate the influence from the non-informative or even harmful sources, we propose a truncated-penalized algorithm with the intuition of dataset clustering. For each study $k\in\{0\}\cup[K]$, with the population parameter $\btheta^*_k$, let $\bTheta = (\btheta_{0},\cdots,\btheta_{K})$, $\hat{\bTheta} = (\hat{\btheta}_{0},\cdots,\hat{\btheta}_{K})\in \reals^{p\times (K+1)}$,
\begin{equation}\label{selection}
	\hat{\bTheta} = \argmin_{\bTheta\in \reals^{p\times (K+1)}}  \frac{1}{N}\sum_{k=0}^{K} \sum_{i=1}^{n_k} \cL_k\left(\bZ_{k,i};\btheta_{k}\right) + \sum_{k=0}^{K} \frac{n_k\lambda_{\cP}}{N} \R\left(\btheta_{k}\right)  +\sum_{k=1}^{K}\frac{n_k \lambda_{\cQ_k}}{N} \cQ\left(\btheta_{k}-\btheta_{0}\right),
\end{equation}
where $N=\sum_{k=0}^{K}n_k$ and $\cQ(\cdot)=\min\left[\cR(\cdot),\tau\right]$  is the truncated norm penalty for some tuning parameter $\tau>0$. Then the first column of $\hat{\bTheta}$, namely $\hat{\btheta}_{0}$, is set to be the estimator of $\btheta_0^*$.

Intuitively, (\ref{selection}) can be viewed as data pooling in a slacker way. Rather than directly enforcing $\btheta_k=\btheta_0$, we penalize the distance between them with penalty function $\cQ$, which possesses the additional truncating nature to cut-off the influence of $\btheta_k$ if it is too far away from $\btheta_0$. As discussed extensively in \cite{duan2023adaptive}, such slackness often leads to robustness. 

Specifically, the tuning parameters $\lambda_{\cQ_k}$ and $\tau$ controls the strength and the range of dataset clustering, respectively. If $\lambda_{\cQ_k}=0$ or $\tau=0$, then (\ref{selection}) calculates each $\hat{\btheta}_{k}$ using the $k$-th dataset alone with penalty $\cR$ and tuning parameter $\lambda_{\cP}$. On the other hand, if $\lambda_{\cQ_k}\rightarrow\infty$ and $\tau\rightarrow \infty$, then optimization problem (\ref{selection}) is equivalent to the optimization problem (\ref{weighted-pooling}) by blind-pooling all the sources and the target dataset. If the individual contrast $\bdelta^*_k=\btheta^*_0-\btheta^*_k$ has certain low-dimensional structure (e.g., sparsity), for some properly chosen $\lambda_{\cQ_k}$ and $\tau$, problem (\ref{selection}) is in the same spirit of the one proposed by \cite{gross2016data,ollier2017regression,li2023estimation} such that the target parameter $\btheta^*_0$ and the contrast vectors $\bdelta^*_k$ from useful sources are estimated simultaneously. On the other hand, if $\bdelta^*_k$ has no such structure but still be small as measured by $\cB_k$, the decomposable regularizer $\cR$ in $\cQ$ tends to penalize $\hat{\btheta}_0-\hat{\btheta}_k$ towards zero. 

In the end, note that (\ref{selection}) possesses the dataset selection capability as those non-informative datasets, identified by $\cR(\hat{\btheta}_0-\hat{\btheta}_k)>\tau$, would have no influence on the estimator $\hat{\btheta}_{0}$ due to the truncation in $\cQ$. Similarly, the estimation of $\hat{\btheta}_{k}$ is also independent from other datasets if $\cR(\hat{\btheta}_0-\hat{\btheta}_k)>\tau$ due to the  same reason. Hence, we also penalize $\btheta_k$ by $\cR(\cdot)$ to ensure the stability in case we have to estimate the high-dimensional vector $\btheta^*_k\in \reals^{p}$ using the $k$-th dataset alone, which is slightly different from the methods in \cite{gross2016data,ollier2017regression,li2023estimation} where only $\btheta_0$ and $\bdelta^*_k$ are penalized.

\subsection{Computation aspects}\label{sec:compute}

To numerically solve problem (\ref{selection}) which is non-convex in nature, we adopt the difference of convex (DC) programming with the alternating direction method of multipliers (ADMM) procedures \citep{boyd2011distributed,wu2016new,fan2019generalized, fan2021shrink}. In this article, we work on two specific statistical models under (\ref{selection}), that is sparse linear regression and generalized low-rank trace regression.

\subsubsection{Sparse linear regression}

We first present the sparse linear regression case, where $\sum_{i\leq n_k}\cL_k\left(\bZ_{k,i};\btheta_{k}\right) = \|\by_{k}-\mathbf{\mathcal{X}}_k\btheta_{k}\|_2^2$ and $\cR = \|\cdot\|_1$. For $\bdelta_k=\btheta_0-\btheta_k$, we focus on the rescaled problem $\hat{\bTheta}=\argmin_{\bTheta,\bdelta} S(\bTheta,\bdelta)$ where
\begin{equation*}\label{Origi_opti}
\begin{aligned}
S(\Theta,\bdelta)=\sum_{k=0}^K \alpha_k \|\by_{k}&-\mathbf{\mathcal{X}}_k\btheta_{k}\|_2^2 +  \sum_{k=0}^K n_k \lambda_{\cP} \|\btheta_{k}\|_1  +\sum_{k=1}^K n_k \lambda_{\cQ_k}\min(\|\bdelta_{k}\|_1,\tau),\\
&\text{subject to}\quad \bdelta_{k}=\btheta_{0}-\btheta_{k},\quad 1\le k\le K.
\end{aligned}
\end{equation*}

We use the difference of convex (DC) programming to convert the non-convex problem into the difference of two convex functions. Specifically, let $S(\bTheta,\bdelta)=S_1(\bTheta,\bdelta)-S_2(\bdelta)$ for $S_1(\bTheta,\bdelta)=\sum_{k=0}^K \alpha_k \|\by_{k}-\mathbf{\mathcal{X}}_k\btheta_{k}\|_2^2 + \sum_{k=0}^K n_k \lambda_{\cP} \|\btheta_{k}\|_1  +\sum_{k=1}^K n_k \lambda_{\cQ_k} \|\bdelta_{k}\|_1$, and $S_2(\bdelta)= \sum_{k=1}^K  n_k \lambda_{\cQ_k}(\|\bdelta_{k}\|_1-\tau)_{+}$, where $a_{+} = \max(a,0)$ for any $a\in \reals$. We replace the convex $S_2(\bdelta)$ by its lower approximation $$S^{(m+1)}_2(\bdelta)=S_2(\hat{\bdelta}^{(m)}) +\sum_{k=1}^{K} n_k \lambda_{\cQ_k} \rbr{\|\bdelta_{k}\|_1-\|\hat{\bdelta}^{(m)}_{k}\|_1} I\rbr{\|\hat{\bdelta}_{k}^{(m)}\|_1\ge\tau},$$
where the superscript $(m)$ represents the $m$-th iteration. Then $S^{(m+1)}(\bTheta,\bdelta)$ can be given as
\begin{equation*}
\begin{aligned}
S^{(m+1)}(\bTheta,\bdelta)&=\sum_{k=0}^K \alpha_k \|\by_{k}-\mathbf{\mathcal{X}}_k\btheta_{k}\|_2^2 +  \sum_{k=0}^K n_k \lambda_{\cP} \|\btheta_{k}\|_1 \\
	&+\sum_{k=1}^K n_k  \lambda_{\cQ_k} \sbr{\|\bdelta_k\|_1
I\rbr{\|\hat{\bdelta}_{k}^{(m)}\|_1<\tau} + \tau I\rbr{\|\hat{\bdelta}_{k}^{(m)}\|_1\ge\tau}},
\end{aligned}
\end{equation*}
which is a convex function of $\bTheta$ and $\bdelta$. We then optimize the upper approximation iteratively:
\begin{equation}\label{eq:DC-slr}
    \rbr{\hat{\bTheta}^{(m+1)},\hat{\bdelta}^{(m+1)}}=\argmin_{\bTheta,\bdelta}\ S^{(m+1)}(\bTheta,\bdelta),\ \text{subject to}\  \bdelta_{k}=\btheta_{0}-\btheta_{k}\ \text{for}\ 1\le k\le K.
\end{equation}

To solve the problem (\ref{eq:DC-slr}), we apply the alternating direction method of multipliers (ADMM) algorithm following \cite{boyd2011distributed} to obtain the global minimizer. These procedures are standard and we leave the implementation details to the appendix for saving space. Finally, the following proposition guarantees the numerical convergence of the DC-ADMM algorithm within finite steps.

\begin{proposition}[Convergence of DC-ADMM] \label{Convergence}
The DC-ADMM algorithm in this section converges in finite steps, namely there exists $m^*<\infty$ that $S(\hat{\bTheta}^{(m)},\hat{\bdelta}^{(m)})=S(\hat{\bTheta}^{(m^*)},\hat{\bdelta}^{(m^*)})$, for $m \geq m^*$. Moreover, $(\hat{\bTheta}^{(m^*)},\hat{\bdelta}^{(m^*)})$ is a Karush-Kuhn-Tucker (KKT) point.
\end{proposition}

\subsubsection{Generalized low-rank trace regression}
 We now introduce the procedure to solve the optimization problem in equation (\ref{selection}) under the case with $\cL_{k}(\bZ_{k,i};\btheta_{k})=-y_{k,i}\eta_{k,i}+b(\eta_{k,i})$ for $\eta_{k,i}=\langle \btheta_k,\bX_{k,i} \rangle$ and $\cR = \|\cdot\|_{\text{N}}$. The difference of convex procedure enables us to focus on a sequence of upper approximations, but optimizing arbitrary loss functions with nuclear norm penalty is still quite challenging. We replace $\bL_{k}(\btheta)=\sum_{1\leq i\leq n_k} \cL_{k}(\bZ_{k,i};\btheta)/n_{k}$ with its quadratic approximation via the iterative Peaceman-Rachford splitting method following \cite{fan2019generalized, fan2021shrink}, simplifying the optimization problem to
\begin{equation*}
\begin{aligned}
\rbr{\hat{\Theta}^{(m+1)},\hat{\bdelta}^{(m+1)},\hat{\bgamma}}&=\argmin_{\Theta,\bdelta,\gamma}\ S^{(m+1)}(\Theta,\bdelta,\bgamma)\\
\text{subject to}&\quad \bdelta_{k}=\btheta_{0}-\btheta_{k},~1\le k\le K,\\
&\quad \bgamma_k=\btheta_k-\hat{\btheta}_k^{(m)},~0\le k\le K,
\end{aligned}
\end{equation*}
for $\bQ(\bgamma_k;\hat{\btheta}^{(m)}_k)=\vec^{\top}\rbr{\bgamma_k}\nabla^2\bL_{k}(\hat{\btheta}^{(m)}_{k})\vec\rbr{\bgamma_k}/2+\vec^{\top}\rbr{\bgamma_k}\vec\sbr{\nabla\bL_{k}\rbr{\hat{\btheta}^{(m)}_{k}}}$ and
\begin{equation}\label{eq:local-problem}
\begin{aligned}
S^{(m+1)}(\bTheta,\bdelta,\bgamma)=&\sum_{k=0}^K n_k\alpha_k
\bQ(\bgamma_k;\hat{\btheta}^{(m)}_k)+\sum_{k=0}^K n_k\lambda_{\cP}\|\btheta_k\|_{\text{N}}\\
&+\sum_{k=1}^{K}n_k\lambda_{\cQ_k}\sbr{\|\bdelta_k\|_{\text{N}}I(\|\hat{\bdelta}_k^{(m)}\|_{\text{N}}<\tau)+\tau I(\|\hat{\bdelta}_k^{(m)}\|_{\text{N}}\ge \tau)}.
\end{aligned}
\end{equation}

Combined with the Theorem 2.1 of \cite{cai2010singular} and \cite{boyd2011distributed}, we can implement standard ADMM procedures to solve the above optimization problem \eqref{eq:local-problem} with nuclear norm penalty. The rest is analogous to the sparse linear regression case, and we also leave the details to the appendix.

\section{Theory}\label{sec:theory}

This section is devoted to the statistical properties of the given method. We begin with the oracle setting where the informative subset $\cA$ is known in advance. For $\cP=\{0\}\cup \cA$, recall that the oracle pooling estimator is
$$\hat{\btheta}_{\cP} =\argmin_{\btheta\in \reals^p} \frac{1}{n_{\cP}}\sum_{k\in \cP}\sum_{i\leq n_k} \cL_k(\bZ_{k,i};\btheta) + \lambda_{\cP} \R(\btheta).$$

We introduce some useful notations from \cite{negahban2012unified}. Define $\bL_{k}(\btheta)=\sum_{1\leq i\leq n_k} \cL_{k}(\bZ_{k,i};\btheta)/n_{k}$, $k=0,\cdots K$, then the optimization problem (\ref{weighted-pooling}) can be rewritten as $\hat{\btheta}_{\cP} = \argmin_{\btheta\in \reals^{p}} \bL_{\cP}(\btheta) + \lambda_{\cP} \cR(\btheta)$, for $\bL_{\cP}=\sum_{k\in\cP}n_k\bL_{k}/n_{\cP}$. Recall that the regularizer $\cR$ is decomposable with respect to $(\cM,\overline{\cM}^{\perp})$ such that $\btheta^*_{0}\in\cM\subset\overline{\cM}$. Let $\R^{*}(\bv)=\sup_{\R(\bu)\leq 1}\langle \bu,\bv\rangle$ be the dual norm of $\R$, define
 $$\delta \bL(\bDelta;\btheta^*_0) = \bL(\btheta^*_0+\bDelta)-\bL(\btheta^*_0) - \left\langle\nabla \bL(\btheta^*_0),\bDelta\right\rangle,$$
 and $\psi(\M)=\sup_{\bu\in \M \backslash \{0\}}\R(\bu)/\|\bu\|$ is the subspace compatibility constant. Throughout this work we shall take the view that $\psi(\M)<\infty$ for simplicity, namely the dimension of $\M$ does not diverge as $p\rightarrow \infty$. In the end, define the cone-like set
\begin{equation*}
\Cb \left(\mathcal{M}, \overline{\mathcal{M}}^{\perp} ; \btheta^{*}\right) :=\left\{\bDelta \in \mathbb{R}^p \mid \mathcal{R}\left(\bDelta_{\overline{\mathcal{M}}^{\perp}}\right)\leq 3 \mathcal{R}\left(\bDelta_{\overline{\mathcal{M}}}\right)+4 \R(\btheta^{*}_{\M ^{\perp}}) \right\},
\end{equation*}
where the subscript, e.g.,  $\bDelta_{\overline{\mathcal{M}}}$ means the projection of $\bDelta$ onto the subspace $\overline{\mathcal{M}}$. In classical fixed $p$ large $n$ regime, the convergence of M-estimators often requires the notion of strong convexity, or equivalently, strictly positive Hessian matrix in a neighbourhood of the population minimizer \citep{van2000asymptotic}. However, in the high-dimensional setting that $p\gg n$, the $p\times p$ Hessian matrix will always be rank-deficient even for linear regression models \citep{negahban2012unified}. Hence, we need to relax strong convexity via restricting the set of directions by the cone-like set $\CC (\cM, \overline{\cM}^{\perp} ; \btheta^*_{0})$. The following proposition follows directly from Theorem 1 of \cite{negahban2012unified}.

\begin{proposition}\label{primal-inform}
   Suppose $\bL_{\cP}$ is convex, differentiable and satisfies the restricted strong convexity (RSC) condition:
\begin{equation}\label{RSC-primal}
	\delta \bL_{\cP}(\bDelta;\btheta^*_{0}) \geq \kappa_{\cP}\|\bDelta\|^2 - \tau_{\cP}, \quad\text{for all }\bDelta\in \CC \left(\cM, \overline{\cM}^{\perp} ; \btheta^*_{0}\right),
\end{equation}
with $\tau_{\cP}\lesssim \lambda_{\cP}\psi^2(\overline{\cM})/\kappa_{\cP}$.  Solve the problem (\ref{weighted-pooling}) with $\lambda_{\cP}\geq 2 \R^{*}\left(\nabla \bL_{\cP}(\btheta^*_{0})\right)$, we have
\begin{equation}\label{primal-bound}
\left\|\hat{\btheta}_{\cP}-\btheta^*_{0}\right\|^2\lesssim  \frac{\lambda^2_{\cP}}{\kappa^2_{\cP}} \psi^2(\overline{\cM}),\quad \cR \left(\hat{\btheta}_{\cP}-\btheta^*_{0}\right)\lesssim \frac{\lambda_{\cP}}{\kappa_{\cP}} \psi^2(\overline{\cM}).
\end{equation}
\end{proposition}

As remarked in \cite{negahban2012unified}, the arguments here are actually deterministic statements about the convex program (\ref{weighted-pooling}). When we deal with particular statistical models, we need to calculate $\R^{*}[\nabla \bL_{\cP}(\btheta^*_{0})]$ and also verify the RSC condition case by case via probabilistic analysis.

Now, we take a closer look at $\lambda_{\cP}$ in the convergence rate of \eqref{primal-bound}. As it is required that $\lambda_{\cP}\geq 2\R^{*}[\nabla \bL_{\cP}(\btheta^*_{0})]$, we focus on the right hand side. Assume that all $\bL_{k}$ have the second-order derivative at $\btheta^{*}_k$, let $\bdelta^*_{k}=\btheta^*_0-\btheta^{*}_k$, by Taylor expansion and triangular inequality,
\begin{equation}\label{var-bias-decom}
	\begin{aligned}
		\cR^{*}\left(\nabla \bL_{\cP}(\btheta^*_0)\right)	&= \cR^{*}\left(\sum_{k\in \cP}n_k \left[\nabla \bL_k(\btheta^{*}_k)+\nabla^2 \bL_k(\btheta^*_0)\bdelta^*_k+ \br_k(\bdelta^*_k)\right] \right)/n_{\cP}\\
		&\leq \underbrace{\cR^{*}\left(\sum_{k\in \cP} n_k \nabla \bL_{k}(\btheta^{*}_k)\right)/n_{\cP}}_{v_{\cP}} + \underbrace{\sum_{k\in \cP} n_k\R^{*}\left( \nabla^2 \bL_k(\btheta^{*}_k)\bdelta^*_k \right)/n_{\cP}}_{b_{\cP}}\\
		&\quad + \underbrace{\cR^{*}\left(\sum_{k\in \cP} n_k\br_k(\bdelta^*_k) \right)/n_{\cP}}_{\text{remainder}}.
	\end{aligned}
\end{equation}

Since $\btheta^{*}_k$ are assumed to be the minimizers of population loss functions, which means $\E[\nabla \bL_k(\btheta^{*}_k)]=0$. The first term $v_{\cP}$ could be viewed as the variance term, and is often well-controlled by standard high-dimensional probabilistic arguments, proportional to $n_{\cP}^{-1/2}$. The second term $b_{\cP}$ could be viewed as the bias term and is the origin of the regularization enlargement. Define the $(\cB_k$,$\cR^*)$-operator norm of the $m\times n$ matrix $\bA$ by $\|\bA\|_{\cB_k \rightarrow\cR^*} = \sup_{\cB_k (\bv)\leq 1} \cR^*\left(\bA \bv\right)$. We have $b_{\cP}\lesssim h$ as long as $\cB_k(\bdelta^*_k)\leq h$ and $\|\nabla^2 \bL_k(\btheta^{*}_k)\|_{\cB_k \rightarrow\cR^*}$ can be bounded above by some constants. Finally, for sufficiently small contrast as measured by $h$, the remainder term could be absorbed into the second term. 


With the previous analysis of the oracle pooling estimator in hand, we are now able to theoretically justify our method of solving the non-convex optimization problem in \eqref{selection}. Indeed, we can show that the oracle pooling estimator $\hat{\btheta}_{\cP}$ is indeed a local minimum of (\ref{selection}) under some mild conditions.

\begin{theorem}[Oracle local minimum]\label{data-set-selection}
	Let $\cA$ denote the informative source datasets, and $\cA^c$ denote the rest. For $k\in \cA$, assume that $\cB_k(\bdelta^*_k)\leq h$ and 
 $$\max(\|\nabla^2 \bL_k(\btheta^*_k)\|_{\cR\rightarrow\cR^*},\|\nabla^2 \bL_k(\btheta^*_k)\|_{\cB_k\rightarrow\cR^*}) \leq M.$$
 In addition, assume that conditions in Proposition \ref{primal-inform} hold for $\cP=\{0\}\cup \cA$.  As for $k\in \cA^c$, define $\hat{\btheta}'_k= \argmin_{\btheta_{k}\in\reals^p}\bL_k\left(\btheta_{k}\right) + \lambda_{\cP} \R\left(\btheta_{k}\right)$, assume that $\cR(\hat{\btheta}'_k-\btheta^*_0)> 2\tau$ for some $\tau>0$ (the same $\tau$ as in $\cQ$). Denote $v_k=\cR^*(\nabla \bL_k(\btheta^*_k))$, for the optimization problem (\ref{selection}) with $\lambda_{\cP}\gtrsim \R^{*}\left(\nabla \bL_{\cP}(\btheta^*_{0})\right)\asymp (v_{\cP}+h)$ and $\lambda_{\cQ_k}\gtrsim  (v_k +h)$, as $\min_{k\in \cA}n_k \rightarrow \infty$, $p\rightarrow \infty$ with $\max_{k\in \cA}v_k<\infty$, $v_{\cP}\rightarrow 0$ and $h\rightarrow 0$, there exists a local minimum $\hat{\bTheta}$ of (\ref{selection}) whose first column satisfies $\hat{\btheta}_0=\hat{\btheta}_{\cP}$.
\end{theorem}

Theorem \ref{data-set-selection}, together with Proposition \ref{Convergence}, justifies the DC-ADMM algorithm's performance from both statistical and computational aspects. In some cases like \texttt{FEDAVG} \citep{mcmahan2017communication}, the oracle pooing estimator $\hat{\btheta}_{\cP}$ is applied to all local clients. That is to say, $\hat{\btheta}_{\cP}$ itself could be directly used as an estimator of $\btheta_0^*$, and its statistical consistency is guaranteed by Proposition \ref{primal-inform}. Meanwhile, for the two-step methods, an additional fine-tuning step using the target dataset is frequently used. Recall that given the oracle pooling estimator $\hat{\btheta}_{\cP}$ (or $\hat{\btheta}_{0}$ from the non-oracle method), we can choose to fine-tune the primal estimator using the target dataset by solving the following problem:
	\begin{equation}\label{lagrangian}
	\hat{\bdelta} = \argmin_{\bdelta\in \reals^p}\frac{1}{n_0} \sum_{i\leq n_0} \cL_0(\bZ_{0,i};\hat{\btheta}_{\cP}+\bdelta) + \lambda_d \R(\bdelta).
	\end{equation}

To some extent, the resulting fine-tuned estimator, denoted as $\hat{\btheta}^{\star}_{\cP}:=\hat{\btheta}_{\cP}+\hat{\bdelta}$, could be viewed as an interpolation between the oracle pooling estimator and the target estimator, which places more emphasis on the personalized aspect of the target dataset. Throughout this work, the additional superscript $\cdot^{\star}$ is used to represent the fine-tuned version of any knowledge transfer estimator. 
 
 When the sources are sufficiently informative, fine-tuning is in fact not statistically necessary, hence we suggest the fine-tuning step to be optional. For instance, in the extreme case that $h=0$, any interpolation towards the target estimator only introduces additional variance to the oracle pooling estimator, leading to statistically sub-optimal performances. This is particularly likely to happen when the target sample size $n_0$ is also relatively small, which is common in practical transfer learning applications. Meanwhile, in certain cases when $\btheta_0^*$ has certain low-dimensional model shift from the sources, as will be shown by numerical simulation and real data analysis in the appendix, the fine-tuning step is able to alleviate such influence in the same way as in \cite{li2020transfer,tian2022transfer} and serves as the final insurance for the satisfying knowledge transfer performance. 

This idea of interpolation between the knowledge transfer estimator and the target estimator for robustness against negative transfer is actually a common practice. Please refer to \cite{duan2023adaptive} where the authors use a similar procedure to (\ref{lagrangian}), so as to avoid the negative impact caused by misusing the data pooling strategy. In the same spirit, \cite{li2023estimation} also propose to split the samples to learn a best linear combination of the knowledge transfer estimator and the individual estimator using target data only, which is proved to be no worse than the single-task estimator with high probability. Cross validation is suggested when choosing the tuning parameter of the fine-tuning step in practice.

\subsection{Specific statistical models}
To end this section, the preceding deterministic arguments are applied on the two specific statistical models, namely sparse linear regression and generalized low-rank trace regression, respectively. Thanks to Theorem \ref{data-set-selection} that guarantees an oracle local minimum of \eqref{selection}, it suffices to discuss the oracle pooling estimator $\hat{\btheta}_{\cP}$ thoroughly in these two specific cases.

\subsubsection{Sparse linear regression}

We assume that the $k$-th dataset consists of $n_k$ {\it i.i.d.} samples from the linear model $y_{k,i} = \langle \btheta^{*}_k, \bX_{k,i} \rangle +\varepsilon_{k,i}$ and $k=0,\dots,K$. Let $\btheta^*_0$ be $s$-sparse with $s<\infty$. To cope with the sparsity structure of $\btheta^*_0$, we take the decomposable regularizer $\cR=\|\cdot\|_1$. We follow \cite{raskutti2010restricted} and (a) assume that the $n_k\times p$ design matrices $\mathbf{\mathcal{X}}_k=\left(\bX_{k,1},\dots,\bX_{k,n_k}\right)^{\top}$ are formed by independently sampling $n_k$ identical $\bX_{k,i}\sim N(0,\bSigma_k)$ with the covariance matrices $\bSigma_k$ satisfying $M_1^{-1}\leq\lambda_{\min}(\bSigma_k)\leq \lambda_{\max}(\bSigma_k)\leq M_1$, which is often referred to as the $\bSigma_k$-Gaussian ensembles; (b) assume that $\varepsilon_{k,i}$ are independently drawn from the same centered sub-Gaussian distribution; and (c) for $k=1,\dots,K$, we assume either $\|\bdelta^*_k\|_1\leq h$ or $\|\bdelta^*_k\|_2\leq h$ (if $\|\bdelta^*_k\|_2\leq h$, we further assume that $p/n_k\rightarrow c_k$ for some positive constant $c_k$) for $\bdelta^*_k=\btheta^*_0-\btheta^{*}_k$.

\begin{corollary}[Sparse linear regression]\label{the-reg}
	Given the above settings, if we solve the problem (\ref{weighted-pooling}) with $\lambda_{\cP}\asymp \left[(\log p/n_{\cP})^{1/2}+h\right]$. As $\min_{k\in \cP}n_k\rightarrow \infty$, $p \rightarrow \infty$ with $\max_{k\in \cP}\log p/n_{k}\rightarrow 0$ and $h\rightarrow 0$, the oracle pooling estimator satisfies
		\begin{equation*}
		\left\|\hat{\btheta}_{\cP}-\btheta^*_0\right\|^2 = O_p\left[\frac{s\log p}{n_{\cP}}+sh^2\right],\quad \left\|\hat{\btheta}_{\cP}-\btheta^*_0\right\|_1 = O_p\left[s\left(\frac{\log p}{n_{\cP}}\right)^{1/2}+sh\right].		
	\end{equation*}
\end{corollary}

In \cite{li2020transfer}, the authors introduce the population oracle pooling parameter as $\btheta^*_{\cP} = \argmin \sum_{k\in \cP} n_k \E [\cL_k(\bZ_k;\btheta)]/n_{\cP}$, where $\cL_k$ is the least squared error loss. In the first step, they first acquire the estimator of $\btheta^*_{\cP}$ as $\tilde{\btheta}_{\cP}$. In the second step, they plug-in $\tilde{\btheta}_{\cP}$ and estimate $\btheta^*_0-\btheta^*_{\cP}$ by solving a fine-tuning optimization problem. When there exists no model shift, namely $h=0$, we have $\btheta^*_{\cP}=\btheta_0^*$ and we shall solve (\ref{weighted-pooling}) with the tuning parameter $\lambda_{\cP} \asymp (\log p/n_{\Pc})^{1/2}$. We also call it as the default rate of regularization, corresponding to the pooled variance term. In fact, in the pooling step of the classical two-step procedures \citep{bastani2021predicting,li2020transfer,tian2022transfer} when $h\neq 0$, the typical rate for regularization is also $\lambda_{\cP} \asymp (\log p/n_{\Pc})^{1/2}$. However, the second fine-tuning step of \cite{li2020transfer} requires heavily on the sparsity of $\btheta^*_{\cP}-\btheta_0^*$, which in turn needs sparse contrast vectors and homogeneous Hessian matrices of the population loss functions. In the case of linear regression, the Hessian matrices are the population covariance matrices of the covariates, namely $\bSigma_k:=\E[\bX_{k,i}\bX_{k,i}^{\top}]$. Hence, these two-step methods are quite sensitive to covariate shift as remarked in \cite{he2024transfusion}. In fact, we now give a toy example where their procedures might fail under almost homogeneous covariates.

\begin{example}[Almost homogeneous covariates]\label{tlad}
	For independent $\bX_{k,i}$ such that $\E (\bX_{k,i})=0$, $\bSigma_k=\E (\bX_{k,i} \bX_{k,i}^{\top})$ and independent noise $\varepsilon_{k,i}\sim N(0,\bI_p)$, let $y_{k,i} = \langle \btheta^{*}_k, \bX_{k,i} \rangle +\varepsilon_{k,i}$,  $k=0,1,2$. In addition, assume that $n_0=n_1=n_2$, $\btheta^*_0$ is $s$-sparse. It is shown in \cite{li2020transfer} that for $\bdelta^*_k=\btheta^*_0-\btheta^{*}_k$, we have
	$$\btheta^*_0-\btheta^*_{\cP} = \left(\sum_{k=0}^{2} \bSigma_k\right)^{-1}\left(\bSigma_1\bdelta_1+\bSigma_2\bdelta_2\right).$$
	
	As for homogeneous covariates such that the covariance matrices are set as $\bSigma_0=\bSigma_1=\bSigma_2$, we have $\btheta^*_0-\btheta^*_{\cP}\asymp(\bdelta_1+\bdelta_2)$, which should be well-bounded by $\|\cdot\|_1$ by triangular inequality once we assume $\|\bdelta^*_k\|_1\leq h$. However, consider the almost-homogeneous covariates such that
	$$\bSigma_0 = \bI_p,\quad \bSigma_1 = \bI_p +  c\bZ_p,\quad \bSigma_2 = \bI_p - c\bZ_p,$$
	where $\bZ_p$ is a fixed realization from the standard Gaussian orthogonal ensemble (GOE), which is the symmetric random matrix with the diagonal elements taken independently from $N(0,2p^{-1})$ and the off-diagonal ones taken independently from $N(0,p^{-1})$. As $p\rightarrow \infty$, the spectrum of $\bZ_p$ is bounded within $[-2,2]$ with high probability, so $\bSigma_1$ and $\bSigma_2$ are positive definite with high probability for sufficiently small $c>0$. Notably, $\|\vec(\bSigma_1)-\vec(\bSigma_2)\|_{\infty}\asymp\|\vec(\bZ_p)\|_{\infty} \lesssim (\log p/p)^{1/2}\rightarrow 0$ as $p\rightarrow \infty$. In this case, we have $\btheta^*_0-\btheta^*_{\cP} \asymp (\bdelta_1+ \bdelta_2) + c\bZ_p(\bdelta_1- \bdelta_2)$. Let $\bdelta_1 = -\bdelta_2 =(h,0,\cdots,0)^{\top}$, $\bZ_p = (\bz_1,\bz_2,\cdots,\bz_p)$, we have
	$$\|\btheta^*_0-\btheta^*_{\cP}\|_1 \asymp  c h\|\bz_1\|_1 \gtrsim  c h\sqrt{p}\rightarrow \infty,$$
	by observing that $\|\bz_1\|_1$ approximately equals to sum of $p$ independent absolute value of $N(0,p^{-1})$.
\end{example}

That is to say, the population pooling parameter $\btheta^*_{\cP}$ could be really far away from $\btheta^*_0$ even if all $\btheta^{*}_k$ are close in terms of $\ell_1$ norm, under heterogeneous covariates. On the other hand, if we instead impose the slightly stronger regularization $\lambda_{\cP} \asymp  (\log p/n_{\Pc})^{1/2}+h$, which could be viewed as the combination of bias and variance, we shall have $\hat{\btheta}_{\cP}$ close to $\btheta^*_0$ as ensured by Corollary \ref{the-reg} above.

We would like to provide some intuitions on such enlarging  regularization. It is well-known that $\cR=\|\cdot\|_1$ is able to element-wise shrink the estimator towards $0$. Recall that $\btheta_0^*$ is supported on some subset $S\subset \{0,1,\dots,p\}$ with cardinality $s$. However, given the potential model shift of sources, it is likely that we are going to have non-zero estimates in $S^c$, whose magnitude is bounded above by $h$. By enlarging the regularization by $h$, we are likely to penalize those estimates in $S^c$ to zero as desired. The price we pay is that additional regularization is at the same time enforced on each element in $S$, which also shrinks more to 0 by $h$, resulting in the additional $sh^2$ term in Corollary \ref{the-reg}. The very same intuition applies to the generalized low-rank trace regression below, except there we use the matrix nuclear norm $\cR=\|\cdot\|_{\text{N}}$ to shrink all the singular values.

In \cite{li2020transfer,li2023estimation}, for their oracle transfer learning estimators to be preferable, they need $h\lesssim (\log p/n_0)^{1/2}$, which results in the rate of $s\log p/n_{\cP}+h^2$. Hence, if $s$ is finite, then the direct oracle pooling estimator via slightly enlarging regularization is no worse than those minimax optimal oracle transfer learning estimators. This is also in the spirit of \cite{chen2021theorem}, where the authors suggest either the target estimator or the oracle pooling estimator is minimax optimal in low-dimensional knowledge transfer problems under certain distance similarity. If $s$ is able to diverge as well, the uniformly over-shrinkage discussed above might be sub-optimal and we suggest considering more sophisticated methods to overcome the covariate shift \citep{li2023estimation,he2024transfusion}. Meanwhile, in many practical cases when $s$ is not too large, we can consider using the more user-friendly $\hat{\btheta}_{\cP}$ acquired by simply enlarging the regularization strength from the default rate.

\subsubsection{Generalized low-rank trace regression}
Then we present the case for generalized low-rank trace regression, the results are analogous to the sparse regression case. Let the $k$-th dataset consist of $n_k$ {\it i.i.d.} samples of $\bZ_{k,i} = (\bX_{k,i},y_{k,i})$, where $\P(y_{k,i}|\bX_{k,i}) \propto \exp\left\{y_{k,i} \eta_{k,i}-b_k(\eta_{k,i}) \right\}$ for $\eta_{k,i}= \langle \btheta^{*}_k, \bX_{k,i} \rangle$. Assume $\btheta^*_0$ is of rank $r$ with $r<\infty$. Recall the subspaces defined in equations (\ref{eq:gltr1}), (\ref{eq:gltr2}) and we take $\cR=\|\cdot\|_{\text{N}}$ as the decomposable regularizer. We follow \cite{fan2019generalized} and assume $\btheta^{*}_k$ to be $d\times d$ square matrices. All analysis could be extended readily to rectangular cases of dimensions $d_1\times d_2$, with the rate replaced by $d=\max(d_1,d_2)$. In addition, assume all datasets share the same $b_k(\cdot)=b(\cdot)$, where $b^{\prime}\left(\eta_{k,i} \right) = \E \left(y_{k,i}|\bX_{k,i}\right)$ is called the inverse link function, and $b^{\prime\prime}\left(\eta_{k,i}\right) = \text{Var} \left(y_{k,i}|\bX_{k,i}\right)$. As also pointed out in \cite{tian2022transfer}, using different $b_k$ is also allowed, but it is less practical.

We take $\bL_k(\btheta)=\sum_{i\leq n_k}\left[-y_{k,i} \langle \btheta, \bX_{k,i} \rangle+b(\langle \btheta, \bX_{k,i} \rangle) \right]/n_k$, whose gradient and Hessian matrices at $\btheta$ are respectively $\nabla \bL_k(\btheta) = \sum_{i\leq n_k}[b^{\prime}(\langle \btheta,\bX_{k,i}\rangle) -y_{k,i}]\bX_{k,i}/n_k$, and $\hat{\bH}_k(\btheta):=	\nabla^2 \bL_k(\btheta) =\sum_{i\leq n_k} b^{\prime \prime}(\langle\btheta, \bX_{k,i}\rangle) \vec(\bX_{k,i}) \vec(\bX_{k,i})^{\top}/n_k$. We make the following assumptions under the guidance of \cite{fan2019generalized}: (a) for each $k\in \cP$, the vectorized version of $\bX_{k,i}$ is taken independently from a sub-Gaussian random vector with bounded $\Psi_2$-norm, namely $\|\vec(\bX_{k,i})\|_{\Psi_2}\leq M_1$; (b) we assume $|b^{\prime\prime}(\eta_{k,i})|\leq M_2$, $|b^{\prime\prime\prime}(\eta_{k,i})|\leq M_3$ almost surely; (c) let $\bH_k(\btheta^{*}_k)=\E \hat{\bH}_k(\btheta^{*}_k)$, assume that $\lambda_{\min}\left[\bH_k(\btheta^{*}_k) \right] \geq \kappa_k$; (d) we assume either $\|\bdelta^*_k\|_{\text{F}}\leq h$, $\|\bdelta^*_k\|_{\text{N}}\leq h$ or $\|\vec(\bdelta^*_k)\|_1\leq h$ for $\bdelta^*_k=\btheta^*_0-\btheta^{*}_k$; and (e) assume that $\|\btheta^*_0\|_{\text{F}}\geq \alpha \sqrt{d}$ for some constant $\alpha$. We claim the following rates of convergence.

\begin{corollary}[Generalized low-rank trace regression]\label{the-glr}
   Given the above settings, if we solve the problem (\ref{weighted-pooling}) with $\lambda_{\cP}\asymp (d/n_{\cP})^{1/2}+h$, as $\min_{k\in \cP}n_k\rightarrow \infty$, $d \rightarrow \infty$ and $h\rightarrow 0$ with $d\lesssim \min_{k\in\cP} n_k$ and $(d/n_{\cP})^{1/2}+h\rightarrow 0$, the oracle pooling estimator satisfies
		\begin{equation*}
		\left\|\hat{\btheta}_{\cP}-\btheta^*_0\right\|_{\text{F}}^2 = O_p\left[\frac{rd}{n_{\cP}}+rh^2\right],\quad \left\|\hat{\btheta}_{\cP}-\btheta^*_0\right\|_{\text{N}} = O_p\left[r\left(\frac{d}{n_{\cP}}\right)^{1/2}+rh\right].		
	\end{equation*}
\end{corollary}

\section{Simulations}\label{sec:simulation}
In this section, we report the numerical performance of the proposed method by simulation study. In the sparse linear regression case, we compare the statistical performance of our truncated-penalized method with some state-of-the-art methods in the literature. As follows, we also numerically validate our claims about enlarging the regularization. Meanwhile, for the generalized low-rank trace regression, to the best of our knowledge, its knowledge transfer problem is still vacant in the literature. Hence, in this case we only compare our algorithm with some ad hoc methods on the shelf.

\subsection{Sparse linear regression}\label{sec:sim-slr}
For the sparse linear regression, we generate data from the linear model $y_{k,i} = \langle \btheta^*_{k}, \bX_{k,i} \rangle +\varepsilon_{k,i}$ for $i=1,\cdots,n_k$ and  $k=0,\cdots,10$, where $\varepsilon_{k,i}$ is drawn independently from $N(0,1)$. For $\bX_{k,i}$, we consider the following cases: (a) homogeneous covariates: draw $\bX_{k,i}$ from $N(\bm{0},\bI_p)$ independently; (b) heterogeneous covariates: let $\bLambda_k$ be a matrix of $1.5p$ rows and $p$ columns whose elements are independently drawn from $ N(0,1)$, then draw  $\bX_{k,i}$ independently from $N(\bm{0},\bSigma_k)$ with $\bSigma_k=2\bLambda_k^{\top}\bLambda_k/(3p)$.

As for $\btheta^*_k$, $k\in\cP:=\{0,\dots,5\}$, we consider two configurations. We set $\btheta^*_{0j}=0.4$ for $j\in[s] = \{1,\cdots,s\}$ where $\btheta^*_{kj}$ represents the $j$-th element of $\btheta^*_k$. Then we consider: (a) sparse contrasts: for each $k\in[5]$, let $H_k$ be a random subset of $[p]$ with $|H_k|=3$ and let $\btheta^*_{kj}=\btheta^*_{0j}-0.4 I\rbr{\cbr{j\in H_k}\cap\cbr{j\neq 1}}$, $\btheta^*_{k1}=-0.4$; (b) dense contrasts: for each $k\in[5]$, let $H_k$ be a random subset of $[p]$ with $|H_k|=p/2$ and let $\btheta^*_{kj}=\btheta^*_{0j}+\xi_j I(j\in H_k)$, where $\xi_j$ is {\it i.i.d.} drawn from $\text{Laplace}(0,0.04)$, $\btheta^*_{k1}=-0.4$. 

\begin{table}[h]
\centering
\caption{The means (standard deviations) of the simulation results for different methods under the settings in this section. For the column ``Datasets", by ``Target" we mean only the target dataset is used, ``Oracle" means we only use the useful datasets, and ``All" means we use all the source datasets. In the case of using all datasets, if the method has dataset selection capability, we report the (TPR,TNR) of dataset selection instead of ``All".}\label{tab:2}
\scalebox{0.7}{
\begin{tabular}{cccccccccc}
\toprule
Estimator&Setting&$\|\cdot-\btheta^*_{0}\|_2$&$\|\cdot^{\star}-\btheta^*_{0}\|_2$&Datasets&Setting&$\|\cdot-\btheta^*_{0}\|_2$&$\|\cdot^{\star}-\btheta^*_{0}\|_2$&Datasets\\
\hline
$\hat{\btheta}_{\text{target}}$&\multirow{7}{*}{$\text{Ho}_s$}&0.738 (0.103)
                                 &NA&Target          &\multirow{7}{*}{$\text{Ho}_d$}&0.747 (0.095)&NA&Target\\
$\hat{\btheta}_{\cP}$&&
                                0.737 (0.029)&0.374 (0.041)&Oracle          &&0.756 (0.036)&0.415 (0.048)&Oracle\\
$\hat{\btheta}_{\cP\cup\cA^c}$&&1.088 (0.078)&0.838 (0.085)&All &                              &1.145 (0.084)&0.901 (0.088)&All \\ 
$\hat{\btheta}_{\text{Agg}}$         &&0.717 (0.078)&0.717 (0.078)&All &                              &0.737 (0.095)&0.737 (0.095)&All \\

$\hat{\btheta}_{\text{CV}}$          &&0.737 (0.029)&0.375 (0.043)&(1.00,1.00)    &                              &0.756 (0.036)&0.414 (0.047)&(1.00,1.00)\\
$\hat{\btheta}_{\text{TF}}$          &&1.138 (0.013)&0.711 (0.068)&Oracle          &                              &1.152 (0.016)&0.723 (0.082)&Oracle\\
$\hat{\btheta}_{\text{TN}}$           &&\bf{0.381 (0.053)}&\bf{0.324 (0.047)}&(1.00,1.00)    &                              &\bf{0.451 (0.078)}&\bf{0.399 (0.080)}&(0.99,1.00)\\ 
\hline
$\hat{\btheta}_{\text{target}}$&\multirow{6}{*}{$\text{He}_s$}&0.787 (0.101)
                                &NA&Target          &\multirow{6}{*}{$\text{He}_d$}&0.788 (0.112)&NA&Target\\

$\hat{\btheta}_{\cP}$&&
                                0.750 (0.032)&0.411 (0.057)&Oracle          &&0.773 (0.032)&0.449 (0.049)&Oracle\\
$\hat{\btheta}_{\cP\cup\cA^c}$&&1.195 (0.091)&0.974 (0.104)&All &                              &1.260 (0.089)&1.048 (0.095)&All \\
$\hat{\btheta}_{\text{Agg}}$         &&0.793 (0.103)&0.793 (0.103)&All &                              &0.790 (0.115)&0.790 (0.115)&All \\
 
$\hat{\btheta}_{\text{CV}}$          &&0.750 (0.032)&0.413 (0.059)&(1.00,1.00)    &                              &0.773 (0.032)&0.451 (0.048)&(1.00,1.00)\\
$\hat{\btheta}_{\text{TF}}$          &&1.141 (0.012)&0.761 (0.080)&Oracle          &                              &1.160 (0.017)&0.765 (0.086)&Oracle\\
$\hat{\btheta}_{\text{TN}}$           &&\bf{0.398 (0.102)}&\bf{0.350 (0.105)}&(0.99,1.00)    &                              &\bf{0.460 (0.050)}&\bf{0.411 (0.044)}&(1.00,1.00)\\ 

\bottomrule
\end{tabular}}
\end{table}

In summary, for the informative sources, we have the following four settings: homogeneous covariates and sparse contrast vectors (with small $\ell_1$ norm), abbreviated as $\text{Ho}_s$; homogeneous covariates and dense contrast vectors (with small $\ell_2$ norm), denoted as $\text{Ho}_d$; heterogeneous covariates and sparse contrast vectors, denoted as $\text{He}_s$; and heterogeneous covariates and dense contrast vectors, denoted as $\text{He}_d$.

As for the non-informative datasets $k\in \cA^c = \{6,\cdots,10\}$, we consider: (a) larger sparse contrasts: for each $k\in \cA^c$, let $H_k$ be a random subset of $[p]$ with $|H_k|=2s$ and let $\btheta^*_{kj}=\btheta^*_{0j}-0.6 I\rbr{\cbr{j\in H_k}\cap\cbr{j\neq 1}}$, $\btheta^*_{k1}=-0.4$; (b) larger dense contrasts: for each $k\in \cA^c$, let $H_k$ be a random subset of $[p]$ with $|H_k|=p/2$ and $\btheta^*_{kj}=\btheta^*_{0j}+\xi_j I(j\in H_k)$, where $\xi_j$ is {\it i.i.d.} from $\text{Laplace}(0,0.2)$, $\btheta^*_{k1}=-0.4$. The reason that we set $\btheta^*_{k1}=-0.4$ for all $k\in[K]$ is to impose a systematic model shift in the first entry, so as to highlight the effect of the fine-tuning step \citep{li2020transfer}.

We compare the following estimators and their fine-tuned versions denoted by the additional superscript $\cdot^{\star}$. To avoid confusions, the proposed estimator by solving \eqref{selection} is denoted as $\hat{\btheta}_{\text{TN}}$ instead of $\hat{\btheta}_{0}$, where TN stands for truncated norm. Our competitor includes the lasso estimator $\hat{\btheta}_{\text{target}}$ on the target dataset, oracle pooling estimator $\hat{\btheta}_{\cP}$ by pooling informative datasets $\cP$, TransFusion estimator $\hat{\btheta}_{\text{TF}}$ by \cite{he2024transfusion}, blind pooling estimator $\hat{\btheta}_{\cP\cup\cA^c}$ by pooling all datasets $\cP\cup\cA^c$, aggregation 
estimator $\hat{\btheta}_{\text{Agg}}$ by \cite{li2023estimation}, CV-based estimator $\hat{\btheta}_{\text{CV}}$ by \cite{tian2022transfer}. We leave the implementation details of these competitors to the appendix for saving space.

We report the performances of these estimators in Table \ref{tab:2} based on 100 replications, where the truncated-penalized algorithm performs quite satisfactorily for its simultaneous estimating the target parameter vector while identifying the useful sources. In addition, the truncated-penalized estimator also outperforms the oracle pooling estimator in some cases, this is due to its superiority of simultaneously estimating the target parameter and the contrasts as also discussed in \citep{gross2016data,ollier2017regression,li2023estimation}. Rigorous theoretical justification for such phenomenon is beyond the scope of this work and left for future pursuits. 
\subsubsection{Enlarging regularization}
Then, we numerically validate our claims about enlarging the regularization. For numerical experiments, we generate useful datasets for $k\in\cP=\{0,\dots 5\}$ in the same way as above, except here we do not deliberately set $\btheta_{k1}=-0.4$ for all $k\in [K]$. Note that $\btheta_{01}=0.4$ and the non-informative sources are not included. 

We consider the following competitors (together with their fine-tuned versions denoted by the additional superscript $\cdot^{\star}$) and sketch how to select tuning parameters for different methods: (a) the target lasso estimator $\hat{\btheta}_{\text{target}}$ using target data and $\lambda_{\text{target}}$ obtained by cross validation; (b) the oracle pooling estimator $\tilde{\btheta}_{\cP}$ targeting the population parameter $\btheta^{*}_{\cP}$, so the tuning parameter $\lambda_{\tilde{\cP}}$ is naturally (and most commonly) chosen by cross validation using all pooling samples; (c) the oracle pooling estimator $\hat{\btheta}_{\cP}$ with the tuning parameter $\lambda_{\hat{\cP}}$ selected by cross validation using target samples.
(d) the fine-tuned versions $\tilde{\btheta}^{\star}_{\cP}$ and $\hat{\btheta}^{\star}_{\cP}$ (for the sake of fairness in comparisons, recall that $\tilde{\btheta}_{\cP}$ is originally used in two-step methods), with the tuning parameters in the second step selected by cross validation using only the target dataset.

\begin{table}[h]
\centering
\caption{The means (standard deviations) of losses under the same settings of Table \ref{tab:2}, except here we do not deliberately set $\btheta_{k1}=-0.4$ for all $k\in [K]$. Note that $\btheta_{01}=0.4$. Here the tuning parameter of $\tilde{\btheta}$ is acquired by cross validation whose validation set consists of the pooled dataset (which is often the default choice). Meanwhile, the tuning parameter of $\hat{\btheta}$ is  acquired by cross validation whose validation set consists of target data only (ending up in a larger $\lambda$ as expected).}\label{tab:1}
\scalebox{0.8}{
\begin{tabular}{cccccccc}
\toprule
Setting&Estimator&$\|\cdot-\btheta^*_{0}\|_2$&$\|\cdot^{\star}-\btheta^*_{0}\|_2$&Selected $\lambda$&$\|\btheta^*_{\cP}-\btheta^*_{0}\|_1$&$\|\hat{\btheta}_{\text{target}}-\btheta^*_{0}\|_2$\\
\hline
\multirow{2}{*}{$\text{Ho}_s$} &$\tilde{\btheta}_{\cP}$  &0.320 (0.026)&0.330 (0.036)&0.038 (0.003)&\multirow{2}{*}{1.038 (0.000)}&\multirow{2}{*}{\centering 0.550 (0.125)}  \\
&$\hat{\btheta}_{\cP}$   &0.322 (0.027)&0.332 (0.037)&0.049 (0.009)\\
\hline
\multirow{2}{*}{$\text{He}_s$} &$\tilde{\btheta}_{\cP}$  &0.356 (0.026)&0.363 (0.033)&0.036 (0.004)&\multirow{2}{*}{5.001 (0.147)}& \multirow{2}{*}{\centering 0.650 (0.197)}  \\
&$\hat{\btheta}_{\cP}$   &0.351 (0.026)&0.358 (0.035)&0.047 (0.007)\\
\hline
\multirow{2}{*}{$\text{Ho}_d$} &$\tilde{\btheta}_{\cP}$  &0.365 (0.035)&0.371 (0.038)&0.038 (0.004)&\multirow{2}{*}{6.565 (0.000)}& \multirow{2}{*}{\centering 0.550 (0.125)}  \\
&$\hat{\btheta}_{\cP}$   &0.348 (0.034)&0.355 (0.036)&0.056 (0.008)\\
\hline
\multirow{2}{*}{$\text{He}_d$} &$\tilde{\btheta}_{\cP}$  &0.389 (0.042)&0.398 (0.049)&0.035 (0.005)&\multirow{2}{*}{8.618 (0.196)}& \multirow{2}{*}{\centering 0.650 (0.197)}  \\
&$\hat{\btheta}_{\cP}$   &0.348 (0.028)&0.359 (0.040)&0.058 (0.009)\\
\bottomrule
\end{tabular}}
\end{table}

We set $p=500$, $n_0=250$, $n_1,\cdots,n_5=400$ and report the results in Table \ref{tab:1} based on 100 replications. While both oracle transfer learning estimators outperform the original target estimation, as $\btheta^*_{\cP}$ gets far away from  $\btheta^*_0$ under the dense contrasts or heterogeneous covariates cases, as indicated by larger $\|\btheta^*_{\cP}-\btheta^*_0\|_1$, the default estimator $\tilde{\btheta}_{\cP}$ turns less reliable. On the other hand, cross validation using only the target samples naturally gives us larger $\lambda$, namely stronger regularization strength. Moreover, it leads to the more reliable estimator $\hat{\btheta}_{\cP}$. In addition, the fine-tuning step seems to be unnecessary in all these cases.

\subsection{Generalized low-rank trace regression}

For generalized low-rank trace regression, we generate datasets for $k=0,\dots,4$ by both identity link $b^{\prime}(x)=x$ (corresponding to linear model) and the logit link $b^{\prime}(x)=1/(1+e^{-x})$ (corresponding to logistic model), with $r=3$, $p_1=p_2=20$, $n_k=400$ for $k\in \{0,\cdots,4\}$. Similarly, $k=0$ is the target, $k=1,2$ are useful, while $k=3,4$ are non-informative. 

\begin{table}[h]
\centering
\caption{The mean and standard error of the simulation results for generalized low-rank trace regression with unknown informative sources. For the column ``Datasets", by ``Target" we mean only the target dataset is used, ``Oracle" means we only use the useful datasets, and ``All" means we use all the source datasets. In the case of using all datasets, if the method include dataset selection capability, we report the (TPR,TNR) of dataset selection instead of ``All".}\label{tab:3}
\scalebox{0.75}{
\begin{tabular}{cccccccc}
\toprule
Estimator&Link&$\|\cdot^{\star}-\btheta^*_{0}\|_{\text{F}}$&Datasets&Estimator&Link&$\|\cdot^{\star}-\btheta^*_{0}\|_{\text{F}}$&Datasets\\
\hline
$\hat{\btheta}_{\text{target}}$ &\multirow{4}{*}{\centering linear}&0.914 (0.073)&All&$\hat{\btheta}_{\text{target}}$&\multirow{4}{*}{\centering logit} &1.473 (0.087)&Target \\
$\hat{\btheta}_{\cP\cup\cA^c}$ &&1.283 (0.085)&All&$\hat{\btheta}_{\cP\cup\cA^c}$& &1.429 (0.062)&All \\
$\hat{\btheta}_{\cP}$& &0.889 (0.056)&Oracle&$\hat{\btheta}_{\cP}$& & \bf{1.282 (0.061)}&Oracle\\

$\hat{\btheta}_{\text{TN}}$ &&\bf{0.581 (0.043)}&(1.00,1.00)&$\hat{\btheta}_{\text{TN}}$ &&1.326 (0.074)&(1.00,0.94)\\
\bottomrule
\end{tabular}}
\end{table}

We report the results in Table \ref{tab:3} based on 100 replications. We draw the conclusion that the truncated-penalized algorithm still performs well in terms of simultaneously identifying the informative auxiliary datasets and recovering low-rank parameters under the current settings.

\section{Real Data Examples}\label{sec:real}
In this section, we show the empirical usefulness of  the proposed method in some real applications. We work on the following two cases concerning the IMDb movie reviews and the air quality in Beijing.

\subsection{IMDb movie reviews}

We first test our algorithm on a publicly available data set of movie reviews from \url{IMDb.com}, which is pre-processed and then used by \cite{gross2016data}. The dataset contains 50K reviews of movies that have been split into a training and test dataset of the same size. For each review there is an integer rating ranging from 1 to 10, where 10 is the best. The dataset only contains positive reviews such that rating $\geq 7$ and negative reviews such that rating $\leq 4$. Following the procedures in \cite{gross2016data}, we first use a binary bag of words representation of the reviews, using only words that were present in at least 500 reviews from the training set, resulting in $p=993$ features.

We focus on the following seven genres of movies. The first three are the most commonly reviewed ones also used in \cite{gross2016data}, namely Drama, Comedy and Horror, with relatively large sample sizes of 4614, 2839 and 1441, respectively. We take these three genres as sources. Then, we also choose four genres, namely Action, Thriller, Sci-Fi and Romance, with relatively small sample sizes of 1002, 655, 223 and 193, respectively, as potential targets. In each experiment, we take one of the latter four genres, say Action, as the target, and treat the other six genres as potential sources of information.

\begin{table}[h]
\centering
\caption{ Mean squared test dataset prediction error $\text{MSE}_{\text{test}}$ and $\text{MSE}^{\star}_{\text{test}}$, where the subscript $\cdot^{\star}$ indicates the fine-tuned version. We take the three most reviewed genres of movies, namely Drama, Comedy and Horror as sources (with relatively large sample sizes of 4614, 2839 and 1441, respectively). Then, we choose four genres, namely Action, Thriller, Sci-Fi and Romance (with relatively small sample sizes of 1002, 655, 223 and 193, respectively), as potential targets. In each experiment, we take one of the latter four genres, say Action, as the target, and treat the other six genres as potential sources.}\label{tab:3}
\scalebox{0.8}{
\begin{tabular}{ccccccccc}
\toprule
Estimator&Target&$\text{MSE}_{\text{test}}$&$\text{MSE}^{\star}_{\text{test}}$&Estimator&Target&$\text{MSE}_{\text{test}}$&$\text{MSE}^{\star}_{\text{test}}$\\
\hline
$\hat{\btheta}_{\text{target}}$&\multirow{5}{*}{Action}&
                               7.949 &NA  &$\hat{\btheta}_{\text{target}}$ &\multirow{5}{*}{Thriller}&11.236 &NA\\
$\hat{\btheta}_{\cP\cup\cA^c}$&&
                                8.032 &8.032  & $\hat{\btheta}_{\cP\cup\cA^c}$  &&8.347 &8.858 \\

$\hat{\btheta}_{\text{CV}}$          &&8.032 &8.026 &                             $\hat{\btheta}_{\text{CV}}$& &8.353 &8.844 \\
$\hat{\btheta}_{\text{TF}}$          &&8.020 &7.747    &$\hat{\btheta}_{\text{TF}}$ &                              &8.316 &8.290 \\
$\hat{\btheta}_{\text{TN}}$           &&\bf{7.423} &\bf{7.423} & $\hat{\btheta}_{\text{TN}}$&                             &\bf{7.386} &\bf{8.061} \\  

\hline
$\hat{\btheta}_{\text{target}}$&\multirow{5}{*}{Sci-Fi}&
                               11.433&NA  &$\hat{\btheta}_{\text{target}}$  &\multirow{5}{*}{Romance}&8.369&NA\\
$\hat{\btheta}_{\cP\cup\cA^c}$&&
                                7.778 &7.778    &$\hat{\btheta}_{\cP\cup\cA^c}$ &&5.672 &6.396\\

$\hat{\btheta}_{\text{CV}}$    &&7.778 &7.778 &$\hat{\btheta}_{\text{CV}}$& &5.657 &6.365 \\
$\hat{\btheta}_{\text{TF}}$          &&7.454 &7.454    &$\hat{\btheta}_{\text{TF}}$ &   &\bf{5.262} &6.248 \\
$\hat{\btheta}_{\text{TN}}$  &&\bf{7.085} &\bf{7.085}&$\hat{\btheta}_{\text{TN}}$ &                              &5.750 &\bf{6.089} \\  
\bottomrule
\end{tabular}}
\end{table}

We compare the same methods as in Section \ref{sec:sim-slr} with the same implementation details, and we report the mean squared prediction error on the test dataset. We report both $\text{MSE}_{\text{test}}$ and $\text{MSE}^{\star}_{\text{test}}$, where the subscript $\cdot^{\star}$ means that we use the fine-tuned estimator for the test dataset prediction. We do not report the aggregation estimator from \cite{li2023estimation} here due to its inherent randomness and instability arising from sample splitting in one single experiment. Its performance is after all not competitive in this particular data case. The results are similar to those in Section \ref{sec:sim-slr}, where our method has satisfying performance across various settings. In the real data case, the truncated-penalized estimator outperforms the other estimators on the test set in three of four cases, except when Romance is the target and the TransFusion estimator has the smallest test dataset mean squared error. 

\subsection{Air pollution data}

Air pollution is an urgent global environmental issue which attracts significant attention from countries worldwide. The majority of the problem is caused by human activities such as industrial emissions and vehicular exhaust. These activities release various pollutants such as particulate matter (PM), nitrogen oxides ($\text{NO}_x$), sulfur dioxide ($\text{SO}_2$), and greenhouse gases into the air. The detrimental effects of air pollution are wide-ranging and severe, which poses severe health risks, environmental degradation, climate change, and the deterioration of ecosystems.

We use transfer learning to analyze the air pollution dataset in Beijing, China. We aim to enhance the next-day prediction performance and provide insights into the winter air pollution problem in Beijing using the proposed method. We collect datasets including Beijing city and eight potential useful cities: Tianjin (TJ), Shijiazhuang (SJZ), Tangshan (TS), Zhangjiakou (ZJK), Hefei (HF), Nanchang (NC), Wuhan (WH) and Shenzhen (SZ). Note that the first four cities are geographically adjacent to Beijing, which might intuitively suffer from similar patterns of air pollution in Beijing.

\begin{figure}[htbp]
	  \centering
	  	    \includegraphics[width=5cm]{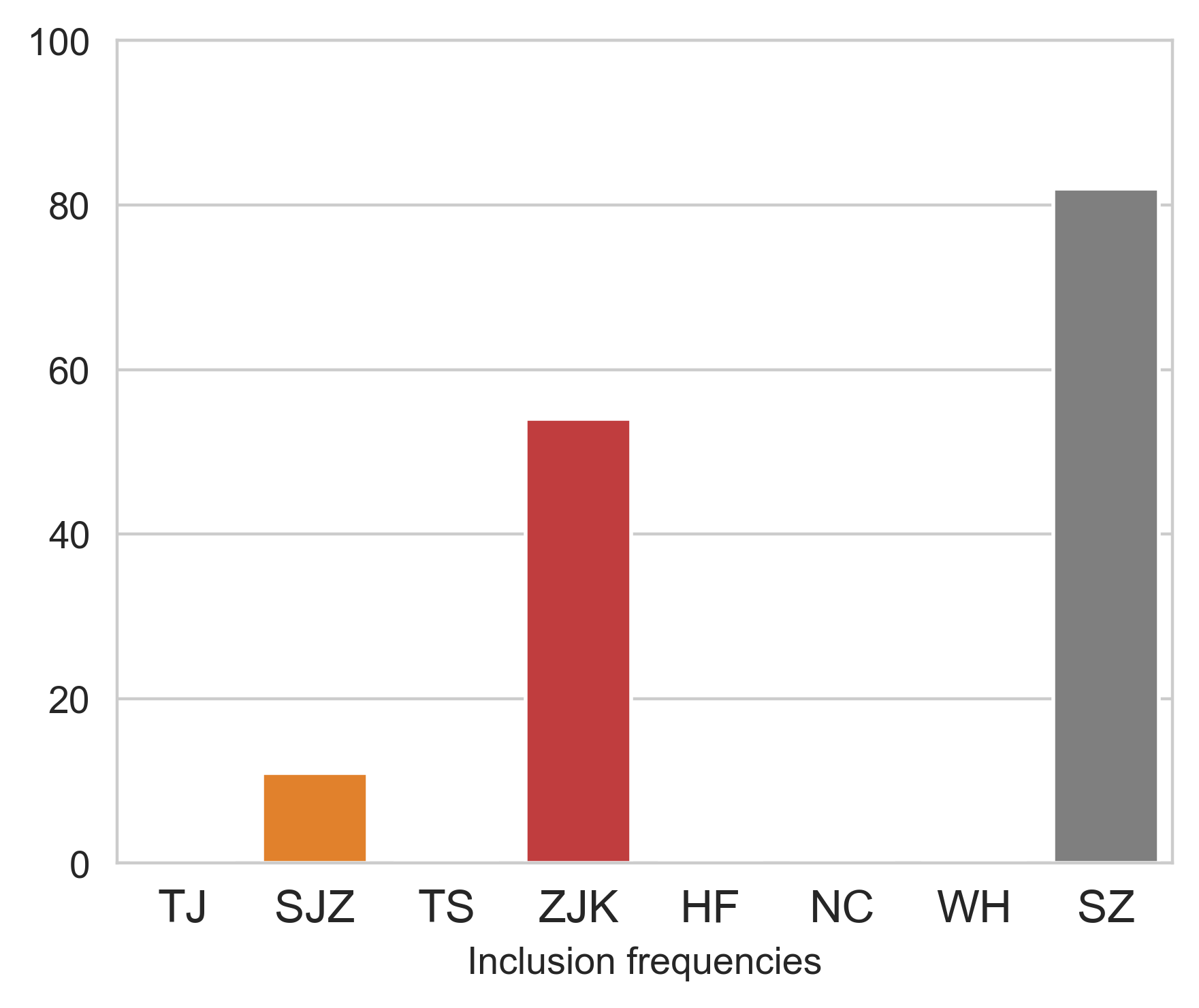}
	  	    \includegraphics[width=5.2cm]{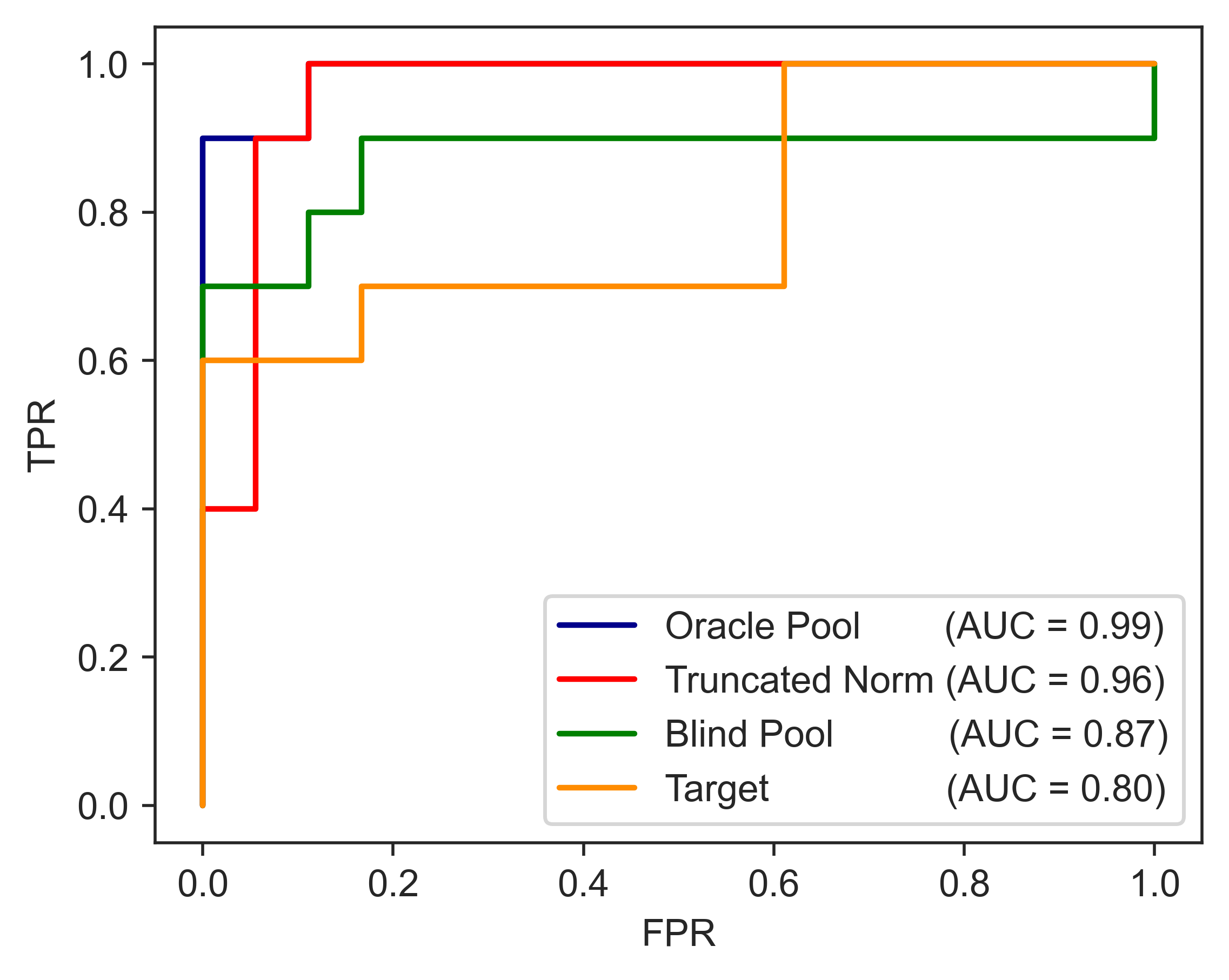}
		\caption{Inclusion frequencies of each source dataset in estimating the daily parameters of the Beijing site using the truncated-penalized algorithm (left). The two most relevant sites (ZJK and SZ) are then used as the oracle informative source datasets in the backtracking rolling windows. The prediction accuracy by various methods are then reported using the receiver operating characteristic (ROC) curve and the area under curve (AUC) metric (right).}\label{fig:real1}
\end{figure}


For each target and source datasets, we collect the daily data from January to February in 2019. In each day $t$, the covariates $\bX_{k,t}$ are matrix-valued data where the rows represent 24 hours and the columns represent the content of six common air pollutants: $\text{PM}_{2.5}$, $\text{PM}_{10}$, $\text{SO}_2$, $\text{NO}_2$, $\text{O}_3$ and CO. The response $y_{k,t}$ is a binary variable, where 1 represents mild pollution while 0 represents relatively good air quality. Clearly, the matrix-valued covariates possess certain column-wise and row-wise correlation, and we add nuclear norm penalty to obtain the low-rank estimation. Specifically, we model the next-day air quality by
$$\P(y_{k,t+1}|\bX_{k,t}) \propto \exp\left\{y_{k,t+1} \langle \btheta^*_k, \bX_{k,t} \rangle-b( \langle \btheta^*_k, \bX_{k,t} \rangle) \right\},$$
 for the logit link $b^{\prime}(x)=1/(1+e^{-x})$, and use the rolling windows approach with window size 31 to forecast the air pollution of February (28 days). For each day, we set $K=8$, $n_k=n_0=31$, $p_1=24$, $p_2=6$, and use respectively the vanilla target estimator, the blind pooling estimator, and the truncated penalized estimator for constructing the next-day prediction.

 We report the inclusion frequencies of each source dataset in estimating the daily parameters of the Beijing site using the truncated-penalized algorithm on the left panel of Figure \ref{fig:real1}, which suggests both Zhangjiakou (ZJK), which is geographically adjacent to Beijing, and Shenzhen (SZ) are informative auxiliary datasets for prediction. Here Shenzhen might have been chosen due to similarities in industrial structure, population, and other social factors with Beijing, which implies that selecting auxiliary datasets based on geographical proximity alone may not be sufficient in the air pollution prediction problems. Then, these two sites (ZJK and SZ) are used as the oracle informative source datasets in the backtracking rolling windows. Note that the information of these sites are useful in most cases is not obtained until the end of the month. We could see that the oracle pooling estimator from backtracking reaches the highest area under curve (AUC) score (AUC=0.99) on the right panel of Figure \ref{fig:real1}, while the truncated-penalized estimator (AUC=0.96) performs comparably.

\section{Discussion}

In this work, we propose to estimate the target parameter vector while simultaneously choosing the informative sources with a non-convex penalty. The proposed algorithm is justified from both statistical and computational aspects. For future work, it is interesting but also challenging to investigate the theoretical properties for the methods that simultaneously estimate the target parameter and the contrasts, see for example \citep{gross2016data,ollier2017regression,li2023estimation}.

\bibliography{main}

\appendix

\section{Additional DC-ADMM Details}
Here we give the additional numerical implementation details concerning the DC-ADMM algorithms under both the sparse linear regression and the generalized low-rank trace regression settings.
\subsection{Sparse linear regression}

To solve the problem (\ref{eq:DC-slr}), the scaled augmented Lagrangian function is set as
$$
\cL_{\rho}(\Theta,\bdelta)=S^{(m+1)}(\bTheta,\bdelta)+\frac{\rho}{2}\sum_{k=1}^{K}||\bdelta_{k}+\btheta_{k}-\btheta_{0}+\bnu_{k}||_2^2-\frac{\rho}{2}\sum_{k=1}^K||\bnu_k||_2^2.
$$
Then the standard ADMM procedure can be implemented as
\begin{equation*}\label{ADMM_Update}
\begin{aligned}
&\hat{\btheta}^{l+1}_k = \argmin_{\btheta_k\in \reals^p}\ \alpha_k \|\by_{k}-\mathbf{\mathcal{X}}_k\btheta_{k}\|_2^2 + n_k \lambda_{\cP} \|\btheta_{k}\|_1+\frac{\rho}{2}\|\hat{\bdelta}_k^l+\btheta_k-\hat{\btheta}_0^l+\hat{\bnu}_k^l\|_2^2,\\
&\hat{\btheta}^{l+1}_0 = \argmin_{\btheta_0 \in \reals^p}\ \alpha_0\|\by_0-\mathbf{\mathcal{X}}_0\btheta_0\|_2^2 + n_0\lambda_{\cP} \|\btheta_0\|_1+\frac{\rho}{2} \sum_{k=1}^{K}\|\hat{\bdelta}_k^l+\hat{\btheta}_k^{l+1}-\btheta_0+\hat{\bnu}_k^l\|_2^2,\\
&\hat{\bdelta}^{l+1}_k = \left\{\begin{array}{ll}
-\hat{\btheta}_k^{l+1}+\hat{\btheta}_0^{l+1}-\hat{\bnu}_k^{l}, & \text{if}~\|\hat{\bdelta}_{k}^{(m)}\|_1\geq \tau,\\
\text{prox}_{n_k\lambda_{\cQ_k}/\rho}(-\hat{\btheta}_k^{l+1}+\hat{\btheta}_0^{l+1}-\hat{\bnu}_k^{l} ),& \text{if}~\|\hat{\bdelta}_{k}^{(m)}\|_1<\tau,\\
\end{array}\right.\\
&\hat{\bnu}_{k}^{l+1}=\hat{\bnu}_{k}^{l}+\hat{\bdelta}_{k}^{l+1}+\hat{\btheta}_{k}^{l+1}-\hat{\btheta}_{0}^{l+1},\\
\end{aligned}
\end{equation*}
where the superscript $l$ denotes the $l$-th step of the ADMM iteration, $\bnu_k$ is the scaled dual variable and the parameter $\rho$ affects the speed of convergence and
the proximal operator under $\ell_1$ penalty could be defined element-wise as $\text{prox}_a(\bb)_i=\rbr{|\bb_i|-a}_{+}\text{sign}(\bb_i)$ \citep{parikh2014proximal}. To acquire $\hat{\btheta}_k^{l+1}$, we construct artificial observations $(\mathbf{\mathcal{X}}'_k,\by'_k)$ and solve  standard lasso problems $\hat{\btheta}^{l+1}_k = \argmin_{\btheta_k\in \reals^p}\|\by'_k-\mathbf{\mathcal{X}}'_k\btheta_{k}\|_2^2+ n_k\lambda_{\cP}\|\btheta_{k}\|_1$ via the cyclic coordinate descent \citep{friedman2010regularization}. Let
\begin{equation*}
\begin{aligned}
(\mathbf{\mathcal{X}}'_0,\by'_0)&=
\begin{pmatrix}
\sqrt{\alpha_0}\mathbf{\mathcal{X}}_0  &  \sqrt{\alpha_0}\by_0\\
\sqrt{\frac{\rho}{2}}\bI_{p\times p}  &  \sqrt{\frac{\rho}{2}}\rbr{\hat{\bdelta}_{1}^{l}+\hat{\btheta}_{1}^{l+1}+\hat{\bnu}_{1}^{l}}\\
\vdots  &  \vdots\\
\sqrt{\frac{\rho}{2}}\bI_{p\times p}  & \sqrt{\frac{\rho}{2}}\rbr{\hat{\bdelta}_{K}^{l}+\hat{\btheta}_{K}^{l+1}+\hat{\bnu}_{K}^{l}}\\
\end{pmatrix}
\end{aligned},
\end{equation*}
while for $k\in [K]$, let
\begin{equation*}
\begin{aligned}
(\mathbf{\mathcal{X}}'_k,\by'_k)&=
\begin{pmatrix}
\sqrt{\alpha_k}\mathbf{\mathcal{X}}_k  & \sqrt{\alpha_k}\by_k\\
-\sqrt{\frac{\rho}{2}}\bI_{p\times p}  & \sqrt{\frac{\rho}{2}}\rbr{\hat{\bdelta}_k^l-\hat{\btheta}_0^l+\hat{\bnu}_k^l}\\
\end{pmatrix}\\
\end{aligned}.
\end{equation*}

For empirical realizations, we set $\hat{\btheta}_k^{0}$ as the lasso solution for the $k$-th datasets and we also set $\hat{\btheta}_k^{(0)}=\hat{\btheta}_k^{0}$, $\hat{\bv}^{0}_{k}=\bm{0}$ for $k\in\cbr{0,1,\cdots,K}$ and~$\hat{\bdelta}_{k}^{(0)}=\hat{\bdelta}_{k}^{0}=\hat{\btheta}_{0}^{(0)}-\hat{\btheta}_{k}^{(0)}$~for~$k\in\cbr{1,2,\cdots,K}$.

\subsection{Generalized low-rank trace regression}

We first present the standard ADMM procedures that can solve \eqref{eq:local-problem} with the nuclear norm penalty. Under the generalized linear model setting, recall that  
$$\nabla\bL_{k}(\btheta_{k})=\sum_{i=1}^{n_k}\sbr{b^{\prime}(\eta_{k,i})-y_{k,i}}\bX_{k,i}/n_k,$$ 
$$\nabla^2\bL_{k}(\btheta_{k})=\sum_{i=1}^{n_k}b^{\prime\prime}(\eta_{k,i})\vec(\bX_{k,i})\vec^{\top}(\bX_{k,i})/n_k.$$

 We define the singular value shrinkage operator $\cS_{\lambda}(\bY)$ for $\bY$ of rank $r$. Let $\bY=\bU\bSigma\bV^{\top}$ where $\bSigma=\diag\cbr{\rbr{\sigma_i}_{1\le i\le r}}$, $\bU$ and $\bV$ are column orthogonal by the singular value decomposition, then $\cS_{\lambda}(\bY):=\bU\bSigma_{\lambda}\bV^{\top}$, $\bSigma_{\lambda}=\diag\cbr{\rbr{\sigma_i-\lambda}_+}$.

the standard ADMM procedure could then be implemented as:

\begin{equation*}
\begin{aligned}
&\hat{\btheta}_{k}^{l+1}=\cS_{\frac{n_k\lambda_{\cP}}{\rho_1+\rho_2}}\rbr{\frac{\rho_1}{\rho_1+\rho_2}\sbr{-\hat{\bdelta}_k^l+\hat{\btheta}_0^l-\hat{\bnu}_k^l}+\frac{\rho_2}{\rho_1+\rho_2}\sbr{\hat{\bgamma}_k^{l}+\hat{\btheta}_k^{(m)}+\hat{\bmu}_k^l}},\quad k=1,2,\cdots,K,\\
&\hat{\btheta}_{0}^{l+1}=\cS_{\frac{n_0\lambda_{\cP}}{K\rho_1+\rho_2}}\rbr{\frac{1}{K\rho_1+\rho_2}\sbr{\rho_1\sum_{k=1}^{K}\rbr{\hat{\bdelta}_k^l+\hat{\btheta}_k^{l+1}+\hat{\bnu}_k^l}+\rho_2\rbr{\hat{\bgamma}_0^{l}+\hat{\btheta}_0^{(m)}+\hat{\bmu}_0^l}}},\\
&\hat{\bdelta}^{l+1}_k =\left\{\begin{array}{ll}
-\hat{\btheta}_k^{l+1}+\hat{\btheta}_0^{l+1}-\hat{\bnu}_k^{l}, & \text{if}~\|\hat{\bdelta}_{k}^{(m)}\|_{\text{N}}\geq \tau,\\
\cS_{n_k\lambda_{\cQ_k}/\rho_1}\rbr{-\hat{\btheta}_{k}^{l+1}+\hat{\btheta}_{0}^{l+1}-\hat{\bnu}_{k}^l}, & \text{if}~\|\hat{\bdelta}_{k}^{(m)}\|_{\text{N}}<\tau,\\
\end{array}\right.\quad k=1,2,\cdots,K,
\end{aligned}
\end{equation*}
\begin{equation*}
\begin{aligned}
& \hat{\bgamma}_k^{l+1}=\argmin_{\bgamma_k}\ n_k\alpha_k
\bQ(\bgamma_k;\hat{\btheta}_k^{(m)})+\frac{\rho_2}{2}\|\bgamma_k-\hat{\btheta}^{l+1}_k+\hat{\btheta}^{(m)}_k+\hat{\bmu}_{k}^{l}\|_{\text{F}}^2,\quad k=0,1,\cdots,K,\\
&\hat{\bnu}_{k}^{l+1}=\hat{\bnu}_{k}^{l}+\hat{\bdelta}_{k}^{l+1}+\hat{\btheta}_{k}^{l+1}-\hat{\btheta}_{0}^{l+1},\quad k=1,2,\cdots,K,\\
&\hat{\bmu}_{k}^{l+1}=\hat{\bmu}_{k}^{l}+\hat{\bgamma}_{k}^{l+1}-\hat{\btheta}_{k}^{l+1}+\hat{\btheta}_{k}^{(m)},\quad k=0,1,\cdots,K,\\
\end{aligned}
\end{equation*}
where $\bnu_k$ and $\bmu_k$ are scaled dual variables and $\rho_1$, $\rho_2$ affect the speed of convergence. By some simple algebra, the updating formula of $\hat{\bgamma}_k^{l+1}$ is
 $$\vec\rbr{\hat{\bgamma}_k^{l+1}}=\Ab^{-}\left\{\rho_2\vec\rbr{\hat{\btheta}_{k}^{l+1}-\hat{\btheta}_{k}^{(m)}-\hat{\bmu}_k^l}-\alpha_k\vec\rbr{\sum_{i=1}^{n_k}\sbr{\rbr{b^{\prime}(\eta_{k,i}^{(m)})-y_{k,i}}\bX_{k,i}}}\right\},$$
 where $\Ab=\sbr{\alpha_k\sum_{i=1}^{n_k}b^{\prime\prime}(\eta_{k,i}^{(m)})\vec(\bX_{k,i})\vec^{\top}(\bX_{k,i})+\rho_2\bI}$ and $\eta_{k,i}^{(m)}=\langle \bX_{k,i},\hat{\btheta}^{(m)}_k\rangle$.

Naturally, we set $\hat{\btheta}_k^{0}$ as the estimator  by \cite{fan2021shrink}, $\hat{\btheta}_k^{(0)}=\hat{\btheta}_k^{0}$, $\hat{\bmu}_k^{0}=\hat{\bgamma}_k^{0}=\bm{0}$, $\hat{\bv}_k^0=\bm{0}$ for $k\in\cbr{0,1,\cdots,K}$ and set ~$\hat{\bdelta}_{k}^{(0)}=\hat{\bdelta}_{k}^{0}=\hat{\btheta}_{0}^{(0)}-\hat{\btheta}_{k}^{(0)}$ hereafter.

Indeed, the fine-tuning step of the generalized low-rank trace regression is also not straightforward due to the non-linearity. To fine-tune the generalized low-rank trace regression with a given $\hat{\btheta}_{\cP}\in\RR^{d_1\times d_2}$, we rewrite the target problem (\ref{lagrangian}) by omitting the subscript $0$ as:
\begin{equation}\label{equ:FineTune1}
\hat{\bdelta}=\argmin_{\bdelta\in\RR^{d_1\times d_2}}\ \frac{1}{n}\sum_{i=1}^{n}\cL(\bZ_{i},\hat{\btheta}_{\cP}+\bdelta)+\lambda_{d}\cR(\bdelta),
\end{equation}
where $\hat{\btheta}_{\cP}$ is known, $\cR=\|\cdot\|_{\text{N}}$ and $\bL(\bdelta)=\sum_{i=1}^{n}\cL\rbr{\bZ_{i},\hat{\btheta}_{\cP}+\bdelta}/n$. Analogously, the local quadratic approximation of (\ref{equ:FineTune1}) is:
\begin{equation*}
\rbr{\hat{\bdelta}^{(m+1)},\hat{\bgamma}}=\argmin_{\bdelta,\bgamma \in\RR^{d_1\times d_2}} \bQ(\bgamma;\hat{\bdelta}^{(m)})+\lambda_{d}\|\bdelta\|_{\text{N}} \quad
\text{subject to}\quad \bgamma=\bdelta-\hat{\bdelta}^{(m)},
\end{equation*}
where for $\eta_i^{(m)} = \langle \bX_{i},\hat{\btheta}_{\cP}+\hat{\bdelta}^{(m)} \rangle$:
$$\bQ(\bgamma;\hat{\bdelta}^{(m)})=\vec^{\top}\rbr{\bgamma}\nabla^2\bL(\hat{\bdelta}^{(m)})\vec\rbr{\bgamma}/2+\vec^{\top}\rbr{\bgamma}\vec\rbr{\nabla\bL(\hat{\bdelta}^{(m)})},$$
$$\nabla\bL(\hat{\bdelta}^{(m)})=\sum_{i=1}^{n}\sbr{b^{\prime}\rbr{\eta_i^{(m)}}-y_i}\bX_i/n,$$
 $$\nabla^2\bL(\hat{\bdelta}^{(m)})=\sum_{i=1}^{n}b^{\prime\prime}\rbr{\eta_i^{(m)}}\vec(\bX_i)\vec^{\top}(\bX_i)/n.$$

Accordingly, we can apply the standard ADMM procedure:

\begin{equation*}
\begin{aligned}
& \hat{\bdelta}^{l+1}=\cS_{\lambda_d/\rho}\rbr{\hat{\bgamma}^{l}+\hat{\bdelta}^{(m)}+\hat{\bnu}^{l}},\\
& \hat{\bgamma}^{l+1}=\argmin_{\bgamma\in \RR^{d_1\times d_2}}\
\bQ(\bgamma;\hat{\bdelta}^{(m)})+\frac{\rho}{2}\|\gamma-\hat{\bdelta}^{l+1}+\hat{\bdelta}^{(m)}+\hat{\bnu}^{l}\|_{\text{F}}^2,\\
&\hat{\bnu}^{l+1}=\hat{\bnu}^{l}+\hat{\bgamma}^{l+1}-\hat{\bdelta}^{l+1}+\hat{\bdelta}^{(m)},\\
\end{aligned}
\end{equation*}
where $\bnu$ is the scaled dual variable and $\rho$ affects the speed of convergence. We have:
 $$\vec\rbr{\hat{\bgamma}^{l+1}}=\Ab^{-}\rbr{\rho\vec\rbr{\hat{\bdelta}^{l+1}-\hat{\bdelta}^{(m)}-\hat{\bnu}^l}-\vec\rbr{\sum_{i=1}^{n}\sbr{\rbr{b^{\prime}\rbr{\eta_i^{(m)}}-y_{i}}\bX_{i}}}/n},$$
where $\Ab=\rho\bI+\sum_{i=1}^{n}b^{\prime\prime}\rbr{\eta_i^{(m)}}\vec(\bX_i)\vec^{\top}(\bX_i)/n$. Note that the ADMM algorithm is not affect by the initial values since the problem is a convex optimization. In practice, we could set $\hat{\bv}^{0}=\bm{0}$, $\hat{\bdelta}^{0}=\hat{\bdelta}^{(0)}=\bm{0}$, $\hat{\bgamma}^{0}=\bm{0}$.

\section{Competitors Implementation Details}

We give the implementation details of the competitors in Section \ref{sec:sim-slr} of the main article. Specifically, for $\hat{\btheta}_{\text{TF}}$, we apply the proposed truncated norm optimization on the datasets $\cP$ with sufficient large $\tau$ to obtain $(\hat{\btheta}_{0},\cdots,\hat{\btheta}_{K})$. Then $\hat{\btheta}_{\text{TF}}$ is defined as the weighted average of $\hat{\btheta}_k$, i.e., $\hat{\btheta}_{\text{TF}}=\sum_{k\in\cP} n_k \hat{\btheta}_k/n_{\cP}$. For $\hat{\btheta}_{\text{Agg}}$, we first randomly divide the target data into two groups $\cT_1$ and $\cT_2$ with equal size, then we obtain the $\btheta_0$ estimator $\hat{\btheta}^1_{\text{Agg}}$ by truncated norm optimization on the datasets $\cT_1\cup [K]$ with sufficient large $\tau$. We can also obtain $\hat{\btheta}^2_{\text{Agg}}$ by lasso estimator on the dataset $\cT_1$. Then for $\bar{\bTheta}=(\hat{\btheta}^1_{\text{Agg}},\hat{\btheta}^2_{\text{Agg}})$, the aggregation estimator is defined as $\hat{\btheta}_{\text{Agg}}=\bar{\bTheta}\bm{\eta}$ where
$$\bm{\eta}=\argmin_{\bm{\eta}\in \text{positive simplex}} \cL_{0}(\Zb_{\cT_2},\bar{\bTheta}\bm{\eta}),\quad \Zb_{\cT_2}\text{ are samples from }\cT_2.$$
For $\hat{\btheta}_{\text{CV}}$, we first estimate $\hat{\cA}$, which is used to obtain the oracle pooling estimator $\hat{\btheta}_{\text{CV}}$. We randomly divide the target data into three groups with equal size, denoted as $\cT_r$, $r\le 3$. Next for each $r$, we obtain the lasso estimator $\hat{\btheta}_{\text{Target}}^r$ on the $\cT_{-r}$, where $\cT_{-r}$ is the target data without $\cT_{r}$. Then we obtain $\hat{\btheta}_{k}^{r}$ by truncated norm optimization on the datasets $k\cup \cT_{-r}$ with sufficient large $\tau$. Accordingly, we can calculate the loss function $\cL_0(\Zb_{\cT_r},\hat{\btheta}_{k}^{r})$ for each $k$ and $r$. Finally, we calculate $\cL_0^k=\sum_{r=1}^3\cL_0(\Zb_{\cT_r},\hat{\btheta}_{k}^{r})/3$, $\cL_0^0=\sum_{r=1}^3\cL_0(\Zb_{\cT_r},\hat{\btheta}_{0}^{r})/3$,
and $\hat{\sigma}=\sqrt{\sum_{r=1}^{3}(\cL_0(\Zb_{\cT_r},\hat{\btheta}_{0}^{r})-\cL_{0}^{0})^2/2}$. Then we have  $\hat{\cA}=\cbr{k\ne 0, \cL_0^k-\cL_0^0\le C_0(\hat{\sigma}\vee0.01)}$. We set $C_0=1$ in the experiments.

\section{Proofs of the Theoretical Results}
Finally, we present the proofs of our theoretical results.

\subsection{Proof of Proposition \ref{Convergence}}
Recall that $\hat{\bTheta}^{(m+1)}$ and $\hat{\bdelta}^{(m+1)}$ would be the minimizer of $S^{(m+1)}(\bTheta,\bdelta)$, then we obtain
\begin{equation}
\begin{aligned}
0&\le S\left(\hat{\bTheta}^{(m)},\hat{\bdelta}^{(m)}\right)=S^{(m+1)}\left(\hat{\bTheta}^{(m)},\hat{\bdelta}^{(m)}\right)\\
&\le S^{(m)}\left(\hat{\bTheta}^{(m)},\hat{\bdelta}^{(m)}\right)\le
S^{(m)}\left(\hat{\bTheta}^{(m-1)},\hat{\bdelta}^{(m-1)}\right)\\
&=S\left(\hat{\bTheta}^{(m-1)},\hat{\bdelta}^{(m-1)}\right).
\end{aligned}
\end{equation}
The remaining parts can be obtained follows similar arguments as those in \cite{wu2016new, liu2023cluster}.

\subsection{Proof of Theorem \ref{data-set-selection}}

Note that the truncated norm penalty makes the problem (\ref{selection}) non-convex, so all minimums here are discussed in a local manner. It helps to decompose (\ref{selection}) into sub-problems. First, for any $\btheta'_{0}$ in some set $\cO_{\cP}$, we acquire the best response of $\hat{\btheta}_k$ as

\begin{equation}\label{fixed_theta0}
\begin{aligned}
	\hat{\btheta}_{k}(\btheta'_{0}) &=  \argmin_{\btheta_{k}\in\reals^p} \hat{\bL}_k(\btheta_{k}; \btheta'_{0}) \\
	&= \argmin_{\btheta_{k}\in\reals^p} \underbrace{\bL_k\left(\btheta_{k}\right) + \lambda_{\cP} \R\left(\btheta_{k}\right)}_{\text{single-task}(k)}+ \underbrace{\lambda_{\cQ_k}\min\left[\cR\left(\btheta_{k}-\btheta'_{0}\right),\tau\right]}_{\text{TNP}(k)},
\end{aligned}
\end{equation}
where TNP stands for the truncated norm penalty. We then plug-in the best responses and locally solve for $\hat{\btheta}_{0}$ by
\begin{equation}\label{op_theta0}
	\hat{\btheta}_{0} = \argmin_{\btheta'_{0}\in \cO_{\cP}} \left[\frac{n_0}{N} \bL_0(\btheta'_{0})+  \frac{n_0}{N} \lambda_{\cP} \R\left(\btheta'_{0}\right)\right]+ \sum_{1\leq k\leq K} \frac{n_k}{N}\hat{\bL}_k\left(\hat{\btheta}_{k}(\btheta'_{0});\btheta'_{0})\right).
\end{equation}

For the informative datasets $k\in\cA$, recall the variance-bias decomposition in (\ref{var-bias-decom}), we have $\cR^{*}\left(\nabla \bL_{\cP}(\btheta^*_0)\right)\lesssim v_{\cP}+h$ since $\|\nabla^2 \bL_k(\btheta^*_k)\|_{\cB_k\rightarrow\cR^*} \leq M$ for $k\in \cA$ and $h\rightarrow 0$. According to Proposition \ref{primal-inform}, for $\lambda_{\cP}\gtrsim v_{\cP}+h$ we have $\cR(\hat{\btheta}_{\cP}-\btheta_0)\lesssim\lambda_{\cP}$ for the oracle pooling estimator $\hat{\btheta}_{\cP}$. Let $l_{\cP}\asymp v_{\cP}+h$ be sufficiently large, so that the open set $\cO_{\cP} = \{\btheta|\cR(\btheta-\btheta^*_0)<l_{\cP}\}$ contains $\hat{\btheta}_{\cP}$. Now, for any $\btheta'_{0}\in \cO_{\cP}$, define $\hat{\bdelta}'_{k}(\btheta'_{0})=\btheta'_{0}-\hat{\btheta}_{k}(\btheta'_{0})$, we rewrite (\ref{fixed_theta0}) in the open set $\cR(\hat{\bdelta}'_{k})<\tau$ as

\begin{equation*}
		\hat{\bdelta}'_{k}(\btheta'_{0}) = \argmin_{\cR(\bdelta_{k})<\tau} \bL_k\left(\btheta'_{0}-\bdelta_{k}\right) + \lambda_{\cP} \R\left(\btheta'_{0}-\bdelta_{k}\right)+ \lambda_{\cQ_k}\cR\left(\bdelta_{k}\right).
\end{equation*}

By the convexity of $\bL_k$ and triangular inequalities we have
\begin{equation}\label{eq:shrinks}
	\begin{aligned}
	 \bL_k\left(\btheta'_{0}-\bdelta_{k}\right) &+ \lambda_{\cP} \R\left(\btheta'_{0}-\bdelta_{k}\right)+ \lambda_{\cQ_k}\cR\left(\bdelta_{k}\right)\\
		&\geq  \bL_k\left(\btheta'_{0}\right)- \cR^*\left(\nabla \bL_k(\btheta'_{0})\right)\cR(\bdelta_{k}) + \lambda_{\cP} \cR(\btheta'_{0})-\lambda_{\cP}\cR(\bdelta_{k}) + \lambda_{\cQ_k} \cR(\bdelta_{k})\\
		 &\geq \bL_k\left(\btheta'_{0}\right) + \lambda_{\cP} \R\left(\btheta'_{0}\right),
	\end{aligned}
\end{equation}
as long as $\lambda_{\cQ_k}\geq \lambda_{\cP}+\cR^*\left(\nabla \bL_k(\btheta'_{0})\right)$. For $v_k=\cR^*(\nabla \bL_k(\btheta^*_k))$, we have
\begin{equation*}
	\begin{aligned}
		\cR^*\left(\nabla \bL_k(\btheta'_{0})\right)&= \cR^*\left(\nabla \bL_k(\btheta^*_k+\btheta'_{0}-\btheta^*_k)\right)\\
		&\lesssim  v_k + \cR^*\left(\nabla^2 \bL_k(\btheta^*_k) (\btheta'_{0}-\btheta^*_k)\right)\\
		&\leq v_k + \R^*\left(\nabla^2 \bL_k(\btheta^*_k) (\btheta'_{0}-\btheta^*_{0})\right)+ \R^*\left(\nabla^2 \bL_k(\btheta^*_k) (\btheta^*_{0}-\btheta^*_{k})\right)\\
		&\lesssim v_k+h,
	\end{aligned}
\end{equation*}
since $\max(\|\nabla^2 \bL_k(\btheta^*_k)\|_{\cR\rightarrow\cR^*},\|\nabla^2 \bL_k(\btheta^*_k)\|_{\cB_k\rightarrow\cR^*}) \leq M$, $\cR(\btheta'_{0}-\btheta^*_0)\lesssim v_p+h \rightarrow 0$, $\cB(\btheta^*_{0}-\btheta^*_{k})\leq h$, and $v_{\cP}\lesssim v_k$. That is to say, for $\lambda_{\cQ_k}\gtrsim v_k +h$, we have $\hat{\bdelta}'_k(\btheta'_{0})=\bf{0}$ for all $\btheta'_{0}\in \cO_{\cP}$ according to (\ref{eq:shrinks}).

Then, for non-informative datasets $k\in \cA^c$, the first part of problem (\ref{fixed_theta0}) is essentially the single-task estimation using the $k$-th dataset only, whose minimizer is denoted by $\hat{\btheta}'_k= \argmin_{\btheta_{k}\in\reals^p}\bL_k\left(\btheta_{k}\right) + \lambda_{\cP} \R\left(\btheta_{k}\right)$. For the non-informative study $k$, since $\cR(\hat{\btheta}'_k-\btheta^*_0)>2\tau$ by assumption, we have $\cR(\hat{\btheta}'_k-\btheta'_{0})> \tau$ for $\btheta'_{0} \in \cO_{\cP}$. The second part of (\ref{fixed_theta0}) is then fixed as $\lambda_{\cQ_k}\tau$ in an open neighborhood of $\hat{\btheta}'_k$, so that $\hat{\btheta}'_k$ is indeed a local minimum of (\ref{fixed_theta0}) and $\hat{\btheta}_{k}(\btheta'_{0})=\hat{\btheta}'_{k}$ for all $\btheta'_{0} \in \cO_{\cP}$.

In the end, plug the best responses into the problem (\ref{op_theta0}), the resulting problem is then equivalent to oracle pooling by (\ref{weighted-pooling}) and the solution is $\hat{\btheta}_{0}=\hat{\btheta}_{\cP}$. The proof is complete.

\subsection{Proof of Corollary \ref{the-reg}}
First, for the restricted strong convexity (RSC) condition, we take the following result from \cite{raskutti2010restricted}, such that there are positive constants $\left(\kappa_{k,1}, \kappa_{k,2}\right)$, depending only on $\bSigma_k$, for $\hat{\bSigma}_k=\mathbf{\mathcal{X}}_k^{\top}\mathbf{\mathcal{X}}_k/n_k$,
\begin{equation}\label{sigmak-ensemble}
\left\langle \bDelta, \hat{\bSigma}_k \bDelta\right\rangle \geq \kappa_{k,1}\|\bDelta\|^2-\kappa_{k,2} \frac{\log p}{n_k}\|\bDelta\|_1^2, \quad \text {for all } \bDelta \in \reals^p.
\end{equation}
with probability greater than $1-c_{k,1} \exp \left(-c_{k,2} n_k\right)$. We then focus on the Hessian matrix of $\bL_{\cP}$, which is $\sum_{k\in\cP}n_k \hat{\bSigma}_k /n_{\cP}$.

\begin{lemma}[RSC Conditions]\label{lem-rsc-reg}
	Under the settings of Corollary \ref{the-reg}, let $\bDelta_{s}$ (or $\bDelta_{s^c}$) be the projection of $\bDelta$ onto the $s$-sparse support (or its complement), there exists a positive constant $\kappa_{\cP}$ such that
	\begin{equation}\label{reg-primal-RSC}
	\frac{1}{n_{\cP}}\left\langle \bDelta, \sum_{k\in\cP}n_k \hat{\bSigma}_k \bDelta\right\rangle \geq \left[\kappa_{\cP}-\sum_{k\in\cP} 16s \kappa_{k,2}\frac{\log p}{n_k} \right]\|\bDelta\|^2, \quad \text {for all }\|\bDelta_{s^c}\|_1\leq 3\|\bDelta_{s}\|_1,
	\end{equation}
	with probability greater than $1-\sum_{k\in\cP}c_{k,1} \exp \left(-c_{k,2} n_k\right)$.
\end{lemma}

Then we give the following result concerning the rate of $\cR^{*}\left(\nabla \bL_{\cP}(\btheta^*_0)\right)$.

\begin{lemma}[Convergence Rates]\label{lem-con-reg}
	Under the settings of Corollary \ref{the-reg}, we have as $\min_{k\in \cP}n_k\rightarrow \infty$, $p \rightarrow \infty$ and $h\rightarrow 0$,
	\begin{equation}
		\cR^{*}\left(\nabla \bL_{\cP}(\btheta^*_0)\right)=O_p(\underbrace{\sqrt{\frac{\log p}{n_{\cP}}}}_{v_{\cP}}+\underbrace{h}_{b_{\cP}}).
	\end{equation}
\end{lemma}

 With the help of Lemma \ref{lem-rsc-reg} and Lemma \ref{lem-con-reg}, while $\max_{k\in \cP}\log p/n_{k}\rightarrow 0$ and $h\rightarrow 0$, Corollary \ref{the-reg} holds according to Proposition \ref{primal-inform}. Then we give the proofs of Lemma \ref{lem-rsc-reg} and Lemma \ref{lem-con-reg}.

 \subsubsection{Proof of Lemma \ref{lem-rsc-reg}}	
It is the straightforward consequences of (\ref{sigmak-ensemble}). For $\|\bDelta_{s^c}\|_1\leq 3\|\bDelta_{s}\|_1$, we have $\|\bDelta\|_1=\|\bDelta_s\|_1+\|\bDelta_{s^c}\|_1\leq 4\|\bDelta_s\|_1\leq 4\sqrt{s}\|\bDelta\|$, then we have (\ref{reg-primal-RSC}) holds with $\kappa_{\cP}:= \sum_{k\in\cP}n_k \kappa_{k,1}/n_{\cP}$ by the union bound of probability.
	
\subsubsection{Proof of Lemma \ref{lem-con-reg}}	
	Controlling $\cR^{*}(\nabla \bL_{k}(\btheta^*_{k}))\asymp \|\mathbf{\mathcal{X}}_k^{\top}\bepsilon_k\|_{\infty}/n_k$ is straightforward by noticing that the maximum of a $p$-dimensional vector with zero mean sub-Gaussian elements, with variance proxies of order $n_k$, is controlled by $(n_k\log p)^{1/2}$ using standard union bound arguments. As for $\cR^{*}\left(\nabla \bL_{\cP}(\btheta^*_0)\right)$, for $h$ sufficiently small, it suffices to bound $v_{\cP}=\cR^{*}\left(\sum_{k\in \cP} n_k\nabla \bL_{k}(\btheta^{*}_k)\right)/n_{\cP}$ and $b_{\cP}=\sum_{k\in \cP}n_k \cR^{*}\left( \nabla^2 \bL_k(\btheta^{*}_k)\bdelta^*_k \right)/n_{\cP}$. First, $v_{\cP}\asymp \|\sum_{k\in \cP} \mathbf{\mathcal{X}}_k^{\top}\bepsilon_k\|_{\infty}/n_{\cP}=O_p(\sqrt{\log p/n_{\cP}})$ by noticing that each element of $\sum_{k\in \cP} \mathbf{\mathcal{X}}_k^{\top}\bepsilon_k$ is the sum of $n_{\cP}$ independent centered random variables and is of order $n_{\cP}^{1/2}$, then we have the result by similar union bound arguments. In the end, for $b_{\cP}=\sum_{k\in \cP}n_k \cR^{*}\left( \nabla^2 \bL_k(\btheta^{*}_k)\bdelta^*_k \right)/n_{\cP}$, we proceed by controlling each term $\|\mathbf{\mathcal{X}}_k^{\top}\mathbf{\mathcal{X}}_k \bdelta^*_k\|_{\infty}/n_{\cP}$. Recall that for $m\times n$ matrix $\bA=(a_{ij})$ and its transpose $\bA^{\top}=\left((\bA^{\top})_1,\cdots,(\bA^{\top})_m\right)$, we have
		\begin{equation*}
		\|\bA\|_{1\rightarrow\infty}  = \sup_{\|v\|_1\leq 1} \|\bA v\|_{\infty}=\max _{(i, j) \in\{1, \ldots, m\} \times\{1, \ldots, n\}}\left|a_{i j}\right|,
		\end{equation*}
		$$\|\bA\|_{2\rightarrow \infty}= \sup_{\|\bv\|_2\leq 1} \|\bA\bv\|_{\infty}=\max _{i=1, \ldots, m}\left\|(\bA^{\top})_i\right\|_{2},$$
	so that $\|\mathbf{\mathcal{X}}_k^{\top}\mathbf{\mathcal{X}}_k\mathbf{\mathcal{X}}_k \bdelta^*_k\|_{\infty} \leq \|\mathbf{\mathcal{X}}_k^{\top}\mathbf{\mathcal{X}}_k\|_{1\rightarrow\infty}\|\bdelta^*_k\|_1$, or $\|\mathbf{\mathcal{X}}_k^{\top}\mathbf{\mathcal{X}}_k \bdelta^*_k\|_{\infty}\leq \|\mathbf{\mathcal{X}}_k^{\top}\mathbf{\mathcal{X}}_k\|_{2\rightarrow\infty}\|\bdelta^*_k\|_2$. Note that by Cauchy-Schwarz inequality, the maximum of $|(\mathbf{\mathcal{X}}_k^{\top}\mathbf{\mathcal{X}}_k)_{ij}|$ is obtained on diagonal, where $(\mathbf{\mathcal{X}}_k^{\top}\mathbf{\mathcal{X}}_k)_{ii}=n_k(\bSigma_k)_{i,i}+O_p(n_k^{1/2})$, so that $\|\mathbf{\mathcal{X}}_k^{\top}\mathbf{\mathcal{X}}_k \bdelta^*_k\|_{\infty}\lesssim n_k h$ with probability tending to 1 using union bound again as $\max_{k\in \cP}\log p/n_{k}\rightarrow 0$ and $\|\bdelta^*_k\|_1\leq h$. On the other hand, $\|\mathbf{\mathcal{X}}_k^{\top}\mathbf{\mathcal{X}}_k\|_{2\rightarrow\infty}\leq \lambda_{\max}(\mathbf{\mathcal{X}}_k^{\top}\mathbf{\mathcal{X}}_k)\leq M_3n_k$ almost surely if $p/n_k\rightarrow c_k$ as $n_k$, $p\rightarrow \infty$ according to \cite{yin1988limit,bai1998no,bai2010spectral} for $\|\bdelta^*_k\|_2\leq h$. The proof is complete.

 \subsection{Proof of Corollary \ref{the-glr}}

 We first verify the RSC conditions under the settings of Corollary \ref{the-glr}. Since $\|\vec(\bX_{k,i})\|_{\Psi_2}\leq M_1$ and $\lambda_{\min}\left[\bH_k(\btheta^{*}_k) \right] \geq \kappa_k$ for $k\in \cP$, according to (6.11) of \cite{fan2019generalized}, with probability $1-\exp(-c_k d)$,
 \begin{equation}\label{fan_rsc_bound}
 	\left\langle \vec(\bDelta),\hat{\bH}_k(\btheta^{*}_k)\vec(\bDelta) \right\rangle \geq \kappa_{k,1}\|\bDelta\|_{\text{F}}^2-\kappa_{k,2}\sqrt{\frac{d}{n_k}}\|\bDelta\|_{\text{N}}^2, \quad \text {for all } \bDelta \in \reals^{d\times d}.
 \end{equation}

 As for the rate of convergence, given $\|\vec(\bX_{k,i})\|_{\Psi_2}\leq M_1$ and $|b^{\prime\prime}(\eta_{k,i})|\leq M_2$ almost surely, according to Lemma 1 of \cite{fan2019generalized}, for $d\lesssim n_k$, as $d\rightarrow \infty$,

 \begin{equation}\label{fan_rate_glr}
 	\left\|\frac{1}{n_k} \sum_{i\leq n_k}\left[b^{\prime}\left(\langle \btheta^{*}_k,\bX_{k,i}\rangle\right)-y_{k,i}\right] \bX_{k,i}\right\|_{\text{op}} =O_p(\sqrt{\frac{d}{n_k}}).
 \end{equation}

 Then, we establish Lemma \ref{lem-rsc-glr} and Lemma \ref{lem-con-glr} based on (\ref{fan_rsc_bound}) and (\ref{fan_rate_glr}).

 \begin{lemma}[RSC Conditions]\label{lem-rsc-glr}
	Under the settings of Corollary \ref{the-glr}, for $\bdelta^*_k=\btheta^*_0-\btheta^{*}_k$, let $\bDelta_{\overline{\mathcal{M}}}$ (or $\bDelta_{\overline{\mathcal{M}}^{\perp}}$) be the projection of $\bDelta$ onto $\overline{\mathcal{M}}$ (or $\overline{\mathcal{M}}^{\perp}$). Denote $\hat{\bH}_{\cP} =\nabla^2 \bL_{\cP} =\sum_{k\in \cP}n_k \hat{\bH}_{k}/n_{\cP}$. There exists a positive constant $\kappa_{\cP}$, as $\min_{k\in \cP}n_k\rightarrow \infty$, $d \rightarrow \infty$ and $h\rightarrow 0$ with $(d/n_{\cP})^{1/2}+h\rightarrow 0$, we have
	\begin{equation}\label{glr-primal-RSC}
	  	\left\langle \vec(\bDelta),\hat{\bH}_{\cP}(\btheta^*_0)\vec(\bDelta) \right\rangle \geq \kappa_{\cP}\|\bDelta\|_{\text{F}}^2, \quad \text {for all } \|\bDelta_{\overline{\mathcal{M}}^{\perp}}\|_{\text{N}} \leq 3\|\bDelta_{\overline{\mathcal{M}}}\|_{\text{N}}.
	\end{equation}
	with probability tending to 1.
\end{lemma}

 \begin{lemma}[Convergence Rates]\label{lem-con-glr}
	Under the settings of Corollary \ref{the-glr}, as $\min_{k\in \cP}n_k\rightarrow \infty$, $d \rightarrow \infty$ and $h\rightarrow 0$ with $d\lesssim \min_{k\in\cP} n_k$, we have
	\begin{equation*}
		\cR^{*}\left(\nabla \bL_{\cP}(\btheta^*_0)\right)=O_p\left[\underbrace{\sqrt{\frac{d}{n_{\cP}}}}_{v_{\cP}}+\underbrace{h}_{b_{\cP}}\right].
	\end{equation*}
\end{lemma}

As the proofs in \cite{negahban2012unified} clarify, we require the RSC condition only on the intersection of $\CC$ with a local ball $\{\|\bDelta\|\leq R\}$, where $R\asymp (d/n_{\cP})^{1/2}+h \rightarrow 0$ is the error radius according to Lemma \ref{lem-con-glr}. Given sufficiently small $R$, we have $\delta \bL_{\cP}(\bDelta;\btheta^*_{0})\gtrsim 	\left\langle \vec(\bDelta),\hat{\bH}_{\cP}(\btheta^*_0)\vec(\bDelta) \right\rangle$, so that the RSC conditions holds according to Lemma \ref{lem-rsc-glr}, and Theorem \ref{the-glr} follows naturally from Proposition \ref{primal-inform}. Finally, we give the proofs of Lemma \ref{lem-rsc-glr} and Lemma \ref{lem-con-glr}.

\subsubsection{Proof of Lemma \ref{lem-rsc-glr}}

	It is the straightforward consequences of (\ref{fan_rsc_bound}). For (\ref{glr-primal-RSC}), we first focus on each term, for $\eta_{k,i}= \langle \btheta^{*}_k, \bX_{k,i} \rangle$,
	\begin{equation}\label{glr-rsc-decom}
	\begin{aligned}
	n_{k}	&\left\langle \vec(\bDelta),\hat{\bH}_k(\btheta^*_0)\vec(\bDelta) \right\rangle =\sum_{i\leq n_k} b^{\prime \prime}\left(\left\langle\btheta^*_0, \bX_{k,i}\right\rangle\right) (\langle\bX_{k,i},\bDelta \rangle)^2\\
	&=\sum_{i\leq n_k} b^{\prime \prime}\left(\left\langle\btheta^{*}_k+\bdelta^*_k, \bX_{k,i}\right\rangle\right) (\langle\bX_{k,i},\bDelta \rangle)^2\\
	&= \sum_{i\leq n_k}\left[ b^{\prime \prime}\left(\eta_{k,i}\right) (\langle\bX_{k,i},\bDelta \rangle)^2 + b^{\prime \prime\prime}(\eta_{k,i})\langle \bX_{k,i}, \bdelta^*_k\rangle (\langle\bX_{k,i},\bDelta \rangle)^2\right.\\
	&\quad \left.+r_{k,i}(\langle \bX_{k,i}, \bdelta^*_k\rangle)(\langle\bX_{k,i},\bDelta \rangle)^2\right].
	\end{aligned}
	\end{equation}
	
	First, notice that both $\|\bdelta^*_k\|_{\text{N}}\leq h$ and $\|\vec(\bdelta^*_k)\|_1\leq h$ implies that $\|\bdelta^*_k\|_{\text{F}}\leq h$. By definition of sub-Gaussian random vectors, we have $\|\left\langle\bX_{k,i},\bdelta^*_k\right\rangle\|_{\Psi_2}\leq M_1h$, so that $\left\langle\bX_{k,i},\bdelta^*_k\right\rangle=o_p(1)$ as $h\rightarrow 0$. The third term in the last line of (\ref{glr-rsc-decom}) hence vanishes and it suffices to control first two terms. We then control the first term directly by (\ref{fan_rsc_bound}), for $\|\bDelta_{\overline{\mathcal{M}}^{\perp}}\|_{\text{N}} \leq 3\|\boldsymbol{\Delta}_{\overline{\mathcal{M}}}\|_{\text{N}}$, with probability greater than $1-\exp(-c_k d)$ we have
	 \begin{equation*}
	 \begin{aligned}
	 \sum_{i\leq n_k} b^{\prime \prime}\left(\eta_{k,i}\right)(\langle\bX_{k,i},\bDelta \rangle)^2&\geq n_k\kappa_{k,1}\|\bDelta\|_{\text{F}}^2-\kappa_{k,2}\sqrt{n_k d}\|\bDelta\|_{\text{N}}^2\\
	 &\geq \left[n_k\kappa_{k,1}-32r \kappa_{k,2}\sqrt{n_k d}\right]\|\bDelta\|_{\text{F}}^2\\
	 &\geq  \left[n_k\kappa_{k,1}-32r \kappa_{k,2}\sqrt{n_{\cP} d}\right]\|\bDelta\|_{\text{F}}^2,
	 \end{aligned}
	 \end{equation*}
	  due to the fact that $\|\bDelta\|_{\text{N}}\leq 4 \|\boldsymbol{\Delta}_{\overline{\mathcal{M}}}\|_{\text{N}} \leq 4\sqrt{2r}\|\bDelta\|_{\text{F}}$. Then, to control the second term of (\ref{glr-rsc-decom}), we have $|b^{\prime \prime\prime}(\eta_{k,i})|\leq M_3$ almost surely by assumption, $|\langle \bX_{k,i}, \bdelta^*_k\rangle|=O_p(h)$ as shown earlier, and $(\langle\bX_{k,i},\bDelta \rangle)^2=O_p(\|\bDelta\|_{\text{F}}^2)$. Combining these results, as $n_{\cP}\rightarrow \infty$, $d \rightarrow \infty$ with $d/n_{\cP}\rightarrow 0$, we have by union bound of probability that
	
	  \begin{equation*}
	  	\left\langle \vec(\bDelta),\hat{\bH}_{\cP}(\btheta^*_0)\vec(\bDelta) \right\rangle \geq \kappa_{\cP}\|\bDelta\|_{\text{F}}^2, \quad \text {for all } \|\bDelta_{\overline{\mathcal{M}}^{\perp}}\|_{\text{N}} \leq 3\|\bDelta_{\overline{\mathcal{M}}}\|_{\text{N}},
	\end{equation*}
	with probability tending to 1, where $\kappa_{\cP} = c_1 \sum_{k\in\cP}n_k\kappa_{k,1}/n_{\cP}$ for some constant $c_1$.

\subsubsection{Proof of Lemma \ref{lem-con-glr}}	

Recall that $\cR^{*}(\nabla \bL_{k}(\btheta^*_{k}))	=\|\sum_{i\leq n_k}\left[b^{\prime}\left(\langle \btheta^*_k,\bX_{k,i}\rangle\right)-y_{k,i}\right] \bX_{k,i}/n_k\|_{\text{op}} =O_p(\sqrt{d/n_k})$ comes directly from (\ref{fan_rate_glr}). We focus on $\cR^{*}\left(\nabla \bL_{\cP}(\btheta^*_0)\right)$ for $h$ sufficiently small. Again it suffices to bound $v_{\cP}=\cR^{*}\left(\sum_{k\in \cP} n_k\nabla \bL_{k}(\btheta^{*}_k)\right)/n_{\cP}$ and $b_{\cP}=\sum_{k\in \cP} n_k \cR^{*}\left( \nabla^2 \bL_k(\btheta^{*}_k)\bdelta^*_k \right)/n_{\cP}$. We could use the standard $\varepsilon$-net argument to control $v_{\cP}$ as in Lemma 1 of \cite{fan2019generalized}, which gives $v_{\cP}=O_p(\sqrt{d/n_{\cP}})$. As for $b_{\cP}$, we control each term
	\begin{equation*}
	 \cR^{*}\left(\nabla^2 \bL_k(\btheta^{*}_k)\bdelta^*_k \right)=\frac{1}{n_k}\left\|\sum_{i\leq n_k} b^{\prime \prime}\left(\left\langle\btheta^{*}_k, \bX_{k,i}\right\rangle\right)\left\langle\bX_{k,i},\bdelta^*_k\right\rangle \bX_{k,i}\right\|_{\text{op}}.
	\end{equation*}
	By definition of $\|\cdot\|_{\text{op}}$ and the standard $\epsilon$-net arguments as in (\ref{fan_rate_glr}), we have
	
	\begin{equation*}
		\begin{aligned}
			\left\|\sum_{i\leq n_k} b^{\prime \prime}\left(\left\langle\btheta^{*}_k, \bX_{k,i}\right\rangle\right)\left\langle\bX_{k,i},\bdelta^*_k\right\rangle \bX_{k,i}\right\|_{\text{op}} &= \sup_{\bu\in \cS^{d-1},\bv\in \cS^{d-1}} \left| \sum_{i\leq n_k} b^{\prime \prime}\left(\left\langle\btheta^{*}_k, \bX_{k,i}\right\rangle\right)\left\langle\bX_{k,i},\bdelta^*_k\right\rangle \bu^{\top}\bX_{k,i}\bv\right|\\
			&\leq \frac{16}{7} \max_{\bu\in \cN^{d},\bv\in \cN^{d}} \left| \sum_{i\leq n_k} b^{\prime \prime}\left(\left\langle\btheta^{*}_k, \bX_{k,i}\right\rangle\right)\left\langle\bX_{k,i},\bdelta^*_k\right\rangle \bu^{\top}\bX_{k,i}\bv\right|,
		\end{aligned}
	\end{equation*}
where $\cS^{d-1}$ is the $(d-1)$-dimensional sphere and $\cN^{d}$ is a $1/4$-net on $\cS^{d-1}$. Then, notice that $|b^{\prime \prime}\left(\left\langle\btheta^{*}_k, \bX_{k,i}\right\rangle\right)|\leq M_2$ by definition, while $\|\left\langle\bX_{k,i},\bdelta^*_k\right\rangle\|_{\Psi_2}\leq M_1h$ and $\|\bu^{\top} \bX_{k,i}\bv \|_{\Psi_2}\leq M_1$ due to the fact that $\|\vec(\bX_{k,i})\|_{\Psi_2}\leq M_1$. Since the product of two sub-Gaussian random variables is sub-exponential, we have for all $\bu\in \cN^{d}$ and $\bv\in \cN^{d}$,
	$$\|\left\langle\bX_{k,i},\bdelta^*_k\right\rangle \bu^{\top}\bX_{k,i}\bv\|_{\Psi_1} \leq \|\left\langle\bX_{k,i},\bdelta^*_k\right\rangle\|_{\Psi_2} \|\bu^{\top} \bX_{k,i}\bv \|_{\Psi_2}\leq M_1^2 h,$$
for the sub-exponential norm $\|\cdot\|_{\Psi_1}$. Hence we are facing the sum of $n_k$ independent sub-exponential random variables with sub-exponential norm (controlling both mean and standard error of the random variables) bounded above by $h$. Then by the union bound over all points on $\cN^d\times \cN^d$ following (6.9) of \cite{fan2019generalized},  we obtain $b_{\cP}=O_p[h+h\max_{k\in\cP}(d/n_k)^{1/2}]$. The proof is then complete by noticing that $d\lesssim \min_{k\in\cP} n_k$ by assumption.

\end{document}